\documentstyle[11pt,emulateapj]{article}
 
\catcode`\@=11
\def\gapprox{\mathrel{\mathpalette\@versim>}}
\def\lapprox{\mathrel{\mathpalette\@versim<}}
\def\@versim#1#2{\lower2.9truept\vbox{\baselineskip0pt\lineskip0.5truept
    \ialign{$\m@th#1\hfil##\hfil$\crcr#2\crcr\sim\crcr}}}
\catcode`\@=12
\lefthead{DYER \& REYNOLDS}
\righthead{BRIGHT RADIO SNRs. III. 3C 397}      

\begin{document}

\submitted{Accepted 6/18/1999 for Astrophysical Journal}
\title{Multifrequency Studies of Bright Radio Supernova 
Remnants. III. X-ray and Radio Observations of 3C 397}
\author{K. K. Dyer, S. P. Reynolds}
\affil{North Carolina State University, Raleigh, NC 27695}
\authoremail{Kristy_Dyer@ncsu.edu}

\begin{abstract}

Radio-bright, presumably young supernova remnants offer the
opportunity of studying strong-shock physics and the nature of the
interaction of ejected material with the surrounding medium.  The
relation between radio and X-ray morphology varies considerably among
supernova remnants, with important implications for the physics of the
emission processes at different wavelengths.  We use Very Large Array (VLA) and Roentgen Satellite (ROSAT)
images of the radio-bright supernova remnant 3C 397 (G41.1--0.3) to
examine the shock structure in both thermal X-ray emission and
nonthermal radio emission.  The unusual rectangular morphology can be
seen in VLA maps at 20 and 6 cm wavelength at a resolution of
$6^{\prime\prime}$, and in ROSAT High-Resolution Imager ({\it HRI}) images.  The X-ray images resemble the radio strongly, except for a small, possibly unresolved
X-ray hot spot near the center.  There is no variation in the X-ray
hardness ratio from ROSAT Position Sensitive Proportional Counter data across the remnant, suggesting
that at least between 0.4 and 2 keV, the interior emission is not
different in character from that in the bright shell regions.  Thus 3C
397 is not a member of the ``thermal composite'' or
``mixed-morphology'' class (\cite{Rho1998} 1998).  The remnant is
unpolarized at 20 cm, and has a mean fractional polarization of $1.5
\pm 0.1 \%$ at 6 cm.  The polarized flux, and polarized fraction, peak
inside the remnant at a location not coincident with either an
internal maximum in total-intensity radio emission, or with the X-ray
hot spot.  Spectral-index maps between 6 and 20 cm do not show any
systematic differences associated with interior emission; there
appears to be no ``plerionic'' or pulsar-driven component in 3C 397 at
least as normally characterized by high polarization and a flat radio
spectrum.  Spectral-index values spread about the mean by about
$\Delta \alpha \sim 0.2$, a result consistent with previous work.  The
steep total-intensity profile off the SW edge of 3C 397 allows an
inference of the upstream electron diffusion coefficient, and implies
a mean free path for electron scattering shorter than in the general
interstellar medium but longer than that similarly inferred for Tycho
and SN 1006 (Achterberg, Blandford \& Reynolds 1994).  A simple analysis based on
the observed X-ray flux gives an estimate of the mean density in 3C
397 of about 4 cm$^{-3}$, which would also be enough to depolarize the
20 cm emission completely, as observed.  The remnant age is then of
order $10^3$ yr, and the current shock velocity about 1600 km
s$^{-1}$.  Finally, we speculate on possible
mechanisms producing the X-ray hot spot.


\end{abstract}


\section{Introduction}
\label{sec:Introduction}

\subsection{X-ray and Radio Comparisons of Supernova Remnants}

Understanding and modeling emission from supernova remnants (SNRs) can
provide a rich mine of astrophysical information.  The physics of
strong shocks has applications beyond SNRs to encompass phenomena as
diverse as the earth's bow shock and extragalactic jets.  It has been
long believed that SNRs play an important, perhaps determining role in
the structure of the interstellar medium.  Recent observations seem to
confirm that SNRs are the source of cosmic rays (\cite{Koyama1995} 1995;
\cite{Tanimori1998} 1998) and may produce gamma-ray emission (\cite{Pohl1998} 
1998).  To examine in more detail the physics of these processes,
and to study the interaction of remnants with their immediate
surroundings, the ideal targets for study would be young remnants, with
high shock velocities, and with high surface brightnesses so that they
are easy to observe, and are likely to be in the well-characterized Sedov
evolutionary phase.

Like 3C391 (Moffett \& Reynolds 1994a, hereafter Paper I)  and W49B (Moffett
\& Reynolds 1994b, hereafter Paper II), subjects of previous study by our group,
3C 397 was selected from a list of the ten highest radio-surface-brightness
shell remnants in the Galaxy at 1 GHz. The characteristics of 3C 397 are
detailed in Table \ref{facts}.

\placetable{facts}

3C 397 has striking characteristics that make it ideal for addressing
questions of supernova-remnant physics.  Its high radio brightness suggests that
it is relatively young, perhaps a transition between the historical
shells and older remnants with radiative shock waves such as the
Cygnus Loop.  
Becker, Markert, and Donahue (1985) found internal polarization they described
as coincident with interior radio emission, implying the possibility
that 3C 397 might harbor a pulsar-driven nebula or ``plerion'' in its
interior.  Einstein observations failed to detect enough
emission to give a good idea of the X-ray morphology, which was
essentially unknown until our observations.  Thus it was completely
unclear whether 3C 397 would show the strong radio--X-ray
morphological correlation characteristic of the historical shells, or
the center-brightened X-rays shown by older but still bright remnants
such as W49B (\cite{II} \& \cite{Pye1984} 1984).  Proper 
classification of
3C 397, and characterization of its X-ray emission as shell-like,
thermal but center-brightened, or plerionic, is clearly critical
to further understanding of this object.

As will be detailed below, we have found that the X-ray emission from
3C 397 is highly spatially correlated with the radio.  The similar
morphology leads to several straightforward questions: To what extent
does the X-ray emission track the edge of the radio emission? How does
the location of the X-ray edge, where electrons are heated, compare to
that of the radio edge, presumably the shock location? Do the edge
profiles have the same shape? Can we detect spectral index variations
within the remnant and how are they correlated with other emission?


\subsection{Previous Observations of 3C 397}

3C 397 (G41.1--0.3) has also been referred to in the literature as
HC26, NRAO597, PKS1905+07 and CTB67.  The source originally identified
as 3C 397 is now known to be two sources, the SNR (3C 397 E) and a
thermal H II region (3C 397 W). Failure by early single-dish
measurements to resolve the two sources complicates flux determinations,
as we discuss later.  We will follow standard usage and will refer to
the SNR alone as 3C 397.  H I absorption measurements place the SNR at
$\geq$6.4 kpc, the tangent point in that direction (\cite
{Caswell1975a} 1975, corrected to a solar Galactocentric distance of
8.5 kpc).  No absorption is seen at negative velocities, locating the
remnant closer than 12.8 kpc.  For our calculations, we will assume a
nominal distance to the SNR of 10 kpc, at which distance $1\arcsec =
0.05$ pc and $1\arcmin = 2.9$ pc.

Observations of the radio recombination lines H 159$\alpha$ and H
159$\beta$ by \cite{Cersosimo1990} (1990) place the H II region at a
near distance of 3.6 kpc or a far distance of 9.3 kpc. H 110$\alpha$
measurements by \cite{Downes1980} (1980) give identical values (after
correction to a solar Galactocentric distance of 8.5 kpc).
\cite{Caswell1975a} (1975) favor the near distance to the HII region,
since it would explain the smaller fractional absorption at higher
velocities they observed.  Relatively strong foreground free-free
absorption in the spectrum of 3C 397 ($\tau_\nu = 1$ at $\nu \sim 30$
MHz; \cite{Kassim1989A} 1989a) could conceivably be due to the outer
fringes of the H II region if it is in front of 3C 397.  As we report
below, there is no morphological evidence for interaction between the
SNR and the H II region, and we shall assume the H II region is a
foreground object.



\begin{figure}
\plotone{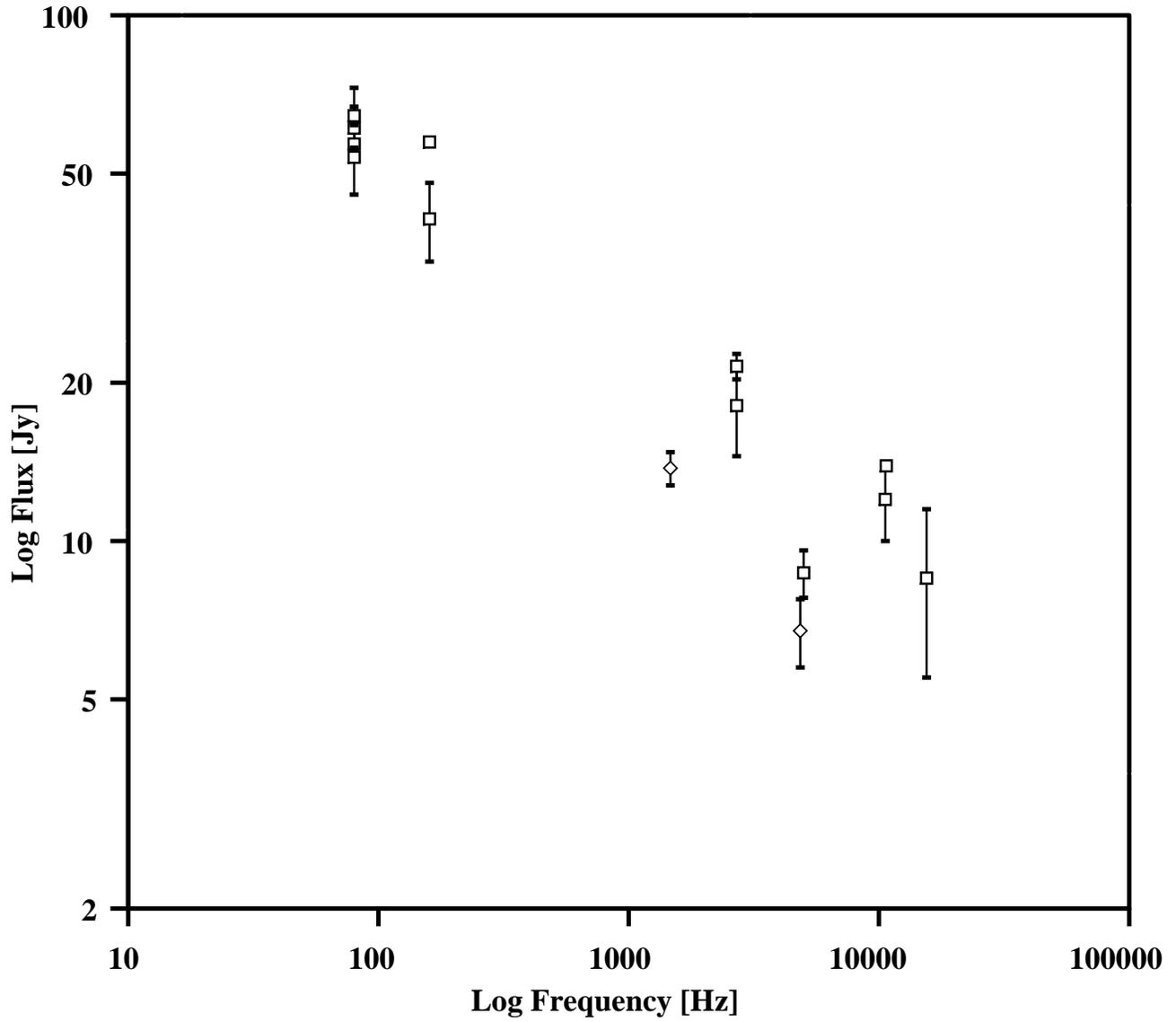}
\caption{Integrated radio spectrum of 3C 397. Diamonds are measurements by Dyer and Reynolds (this paper). Squares indicate measurements by instruments capable of resolving the SNR and HII region. Data compiled by Ford 1994.\label{ford}}
\end{figure}

The total flux density of 3C 397 at centimeter wavelengths is quite
uncertain (see Fig.~\ref{ford} for a compilation from the literature).
Recent 330 MHz observations by \cite{Kassim1992} (1992) found a flux
of 38 Jy for 3C 397, consistent with his earlier study at 30.9 MHz
(\cite{Kassim1989B} 1989b) which reported a spectral fit to his and
earlier data of a power-law with index $\alpha = -0.4$ ($S_\nu \propto
\nu^\alpha$) and a low-frequency exponential turnover due to
foreground free-free absorption, with optical depth unity around 30
MHz.  This fit gives a nominal 1 GHz flux of 27 Jy.
\cite{Green1998}'s (1998) catalogue lists a flux of 22 Jy at 1 GHz,
with a spectral index of $-0.48$.  Green's estimated fluxes are then
18.3 and 10.3 Jy at 20 and 6 cm, respectively, while Kassim's fit
gives 23 and 14 Jy.  In inspecting the original literature reporting
single-dish fluxes for 3C 397, we found that many reports involved
instruments with insufficient angular resolution to separate the H II
region from the SNR.  Furthermore, 3C 397 sits in a region of high
Galactic background.  The 1.4 GHz Bonn survey (Reich, Reich \& F\"{u}rst 1990)
found and removed a smoothly varying background of 54 mJy
arcmin$^{-2}$ near 3C 397.  A smaller plateau of brightness about 19
mJy arcmin$^{-2}$ was not removed, and is visible on the survey maps in
the vicinity of the remnant.  We conclude that the spectrum of 3C 397
is so poorly known at this time that the spectral index is uncertain
by several tenths, and the total fluxes at 20 and 6 cm by factors of
at least 2.  This lack of agreement will hamper our attempts to
interpret spectral index information below.


Becker, Markert \& Donahue (1985) observed 3C 397 with the National Radio Astronomy Observatory's (NRAO) Very Large Array\footnote{The National Radio Astronomy Observatory is
a facility of the National Science Foundation operated under a
cooperative agreement by Associated Universities, Inc.} (VLA) at 6 and 20
cm with a resolution of 14\arcsec, to get a quick idea of its radio
morphology.  They first described its box-like shape, partial shell,
and significant interior emission.  They detected some polarization at
6 cm which they described as coincident with the peak of the interior
total intensity.  A detailed spectral-index study of 3C 397 was
carried out by \cite{AR1993} (1993), also with the VLA at 20 and 6 cm
and a resolution of 14\arcsec. They reported spectral-index variations
of order $\Delta \alpha \sim 0.1$ across the remnant, and found, in
keeping with other young remnants, that these variations do not
coincide with total intensity features.  They concluded that spectral
variations must be determined by different mechanisms than brightness
variations, and found no obvious explanation.

In the first Einstein Slew Survey
(\cite{Elvis1992} 1992), 3C 397 had an Imageing Proportional Counter ({\it IPC}) count rate of 0.45 cts s$^{-1}$
in an exposure of 14 seconds.  \cite{Becker1985} (1985) observed 3C 397
with the Einstein High-Resolution Imager ({\it HRI}), {\it IPC}, and Monitor Proportional Counter ({\it MPC}), detecting 200 counts in a 3200-second
{\it HRI} exposure and 480 counts in an 800-second {\it IPC} exposure.
The {\it HRI} image had low signal to
noise but both the {\it IPC} and the {\it HRI} images indicated some emission
associated with the interior of the radio shell. Thermal fits to
the {\it IPC} and {\it MPC} data did not agree on temperature or column density. 


No emission is apparent at the location of the SNR in the Palomar sky
survey, or in the Infrared Astronomical Satellite (IRAS) catalog of supernova remnants by
Saken, Fesen \& Shull (1992). Lack of optical detection is not surprising
given 3C 397's location in the Galactic plane and distance of
order 10 kpc.

\section{Observations}
\label{sec:Observations}

\subsection{Radio Observations \& Method}
\placetable{observations}

We observed 3C 397 with the VLA
at wavelengths of 6 and 20 cm (4847 MHz and 1468 MHz) in B, C and D
configurations between 1990 April and 1992 January as described in
Table \ref{observations}. Reduced bandwidths of 6250 kHz were used at
20 cm to reduce bandwidth depolarization, with the two
intermediate frequency (IF) channels 1 and 2 separated by 50 MHz in all cases. We observed 3C
286 (1328+307) as our primary flux calibrator and 1821+107 as the
secondary (phase) calibrator and to remove instrumental
polarization. The fluxes for 3C 286 were calculated using new VLA
coefficients: at 1446 MHz, 14.7417 Jy; at 1496 MHz,
14.5014 Jy; at 4836 MHz, 7.5419 Jy; and at 4886 MHz, 7.5903 Jy.
The data were calibrated and edited using NRAO's Astronomical Image
Processing System (AIPS). Visibility data at each frequency were
combined to provide projected interferometer baseline coverage from
0.03 to 3.4 km (400--52000 $\lambda$) at 6 cm and 0.03 to 11.4 km
(200--5300 $\lambda$) at 20 cm. 

We mapped the data with the AIPS task
MX with uniform weighting and deconvolved images using both the CLEAN
algorithm (with a taper of 35k$\lambda$ at 6 and 20 cm to improve the
shape of the dirty beam but without adding zero spacing flux) and
VTESS, a maximum entropy method.  VTESS super-resolves features with
high signal-to-noise; we convolved the VTESS image with the CLEAN
beam before analysis.  
The images were corrected for the
primary-beam response of the individual VLA antennas. We used the
Stokes $V$ polarization images to gauge intrinsic sensitivity, since
no circular polarization beyond ``beam squint'' was expected. The
final sensitivities achieved at each band are quoted in Table
\ref{summary}. An rms of about five times the theoretical thermal noise was
achieved in the CLEAN maps (integration time was $\sim$ 60 minutes with a bandwidth of 25 MHz at 6 cm and 6.25 MHz at 20 cm, see Table \ref{observations}). As expected for the galactic plane the images were confusion limited. For VTESS, arbitrary values of noise can
be specified, but the algorithm was unable to converge for values
lower than about three times the theoretical noise.
 As discussed in Paper I, \S 2 the difficulties of
deconvolving low-surface-brightness extended sources like SNRs often
result in noise levels considerably higher than theoretical. 
Table \ref{noise} compares the theoretical and measured noise.

\placetable{noise}

We compared the CLEAN and maximum entropy deconvolutions as a guide to the
reality of faint features. Those we describe are present in images from
both methods and are therefore likely to be real.  

\subsection{X-ray Observations}

3C 397 was observed by the X-ray Roentgen Satellite (ROSAT) with the
Position-Sensitive Proportional Counter ({\it PSPC}) on 1992 September 28,
and with the High-Resolution Imager ({\it HRI}) between 1994 October 14 and 21. Information on these observations is given in Table \ref{rosat}.

The IRAF/PROS software was used, supplemented with XSPEC and FTOOLS
4.1, to analyze the X-ray data.  Because of the small size of the
remnant compared to the ROSAT telescope field of view, vignetting
corrections were not necessary.  {\it PSPC} spectral data were obtained from
a circular region of radius 170\arcsec, with background determined
from a concentric annulus of inner and outer radii 170\arcsec~
and 338\arcsec.  The {\it HRI} image was smoothed with a Gaussian of
full width at half max (FWHM) 6\arcsec, since the relatively low X-ray surface brightness resulted in the point-source response not being fully sampled over
most of the remnant.

\section{Results}
\label{sec:Results}

        \subsection{Radio}

	\subsubsection{Total Intensity Images}
\begin{figure}
\plotone{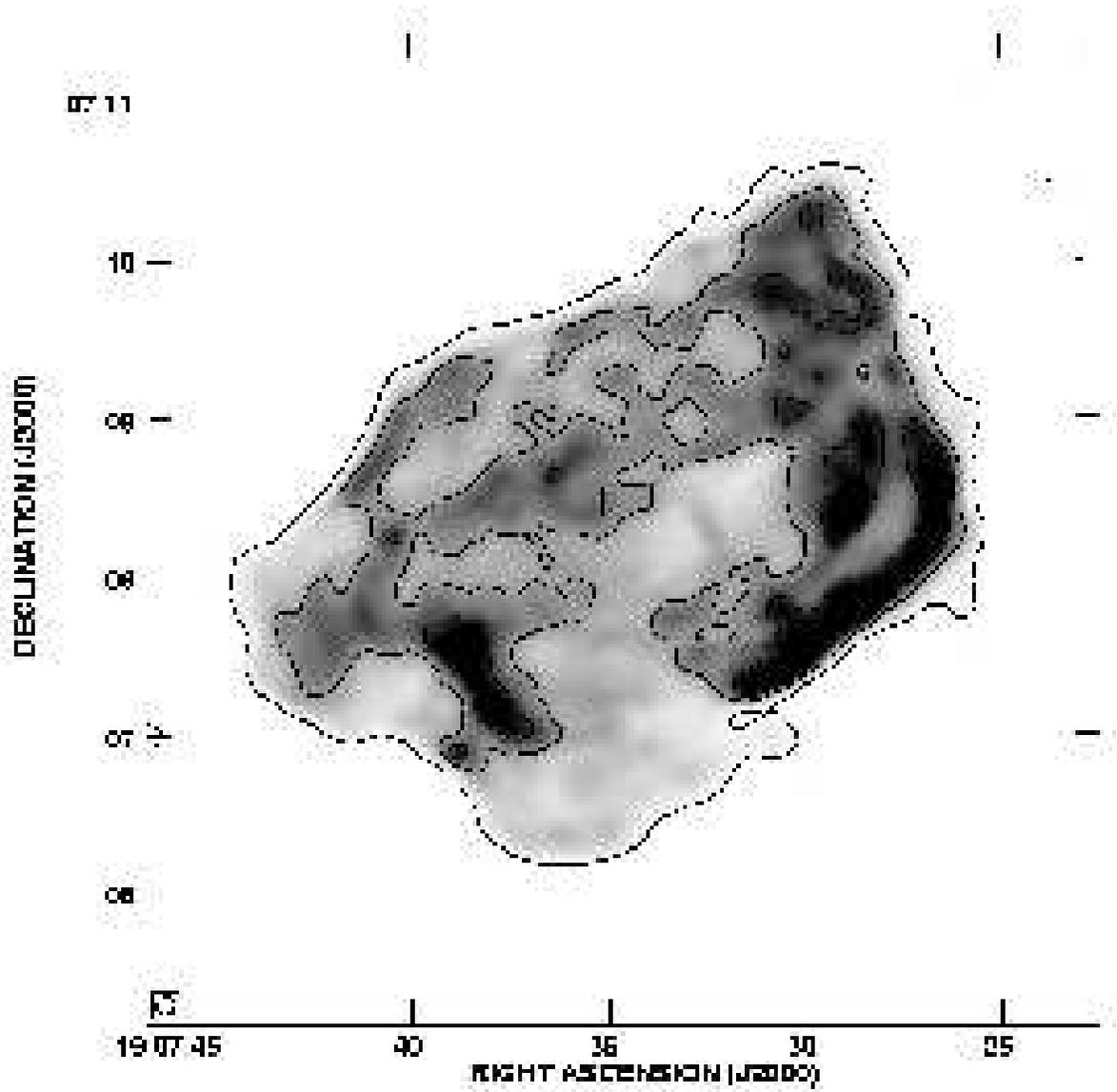}
\caption{20 cm radio image of 3C 397 taken at the VLA. Processed using
maximum entropy method. Barry Clark CLEANed image had no significant
difference. Contours are at 3, 25, 47$\sigma$ where $\sigma$ is the rms noise, 
0.59 mJy beam$^{-1}$. The synthesized beam is 6.9 x 6.6\arcsec. Primary beam 
corrected. The cross indicates the location of the X-ray hot spot $\alpha$ 
19$^h$ 7$^m$ 35\fs13, $\delta$ 7\arcdeg~8\arcmin~28\farcs76.\label{lvtc}}    
\end{figure}

\begin{figure}
\plotone{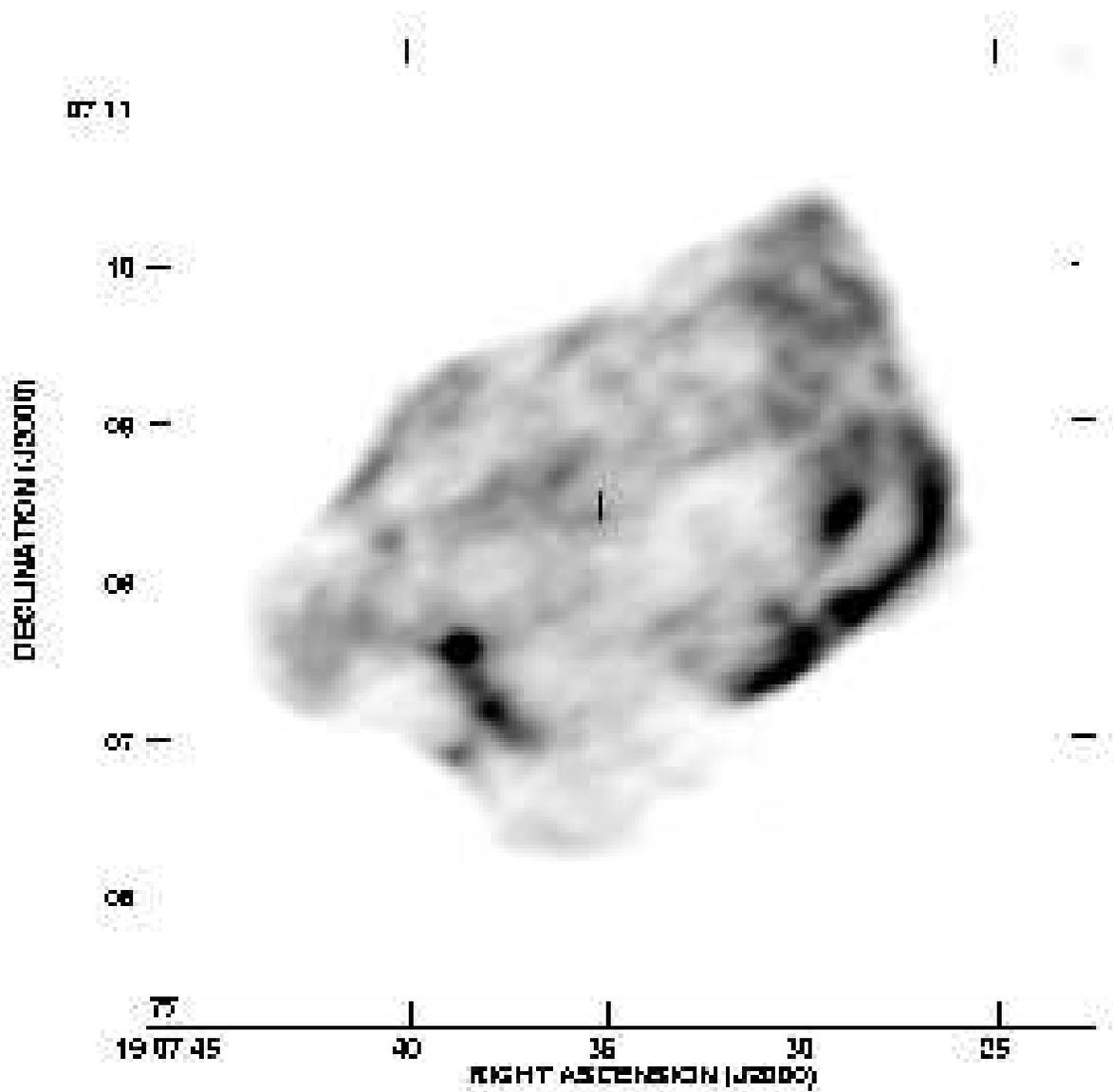}
\caption{6 cm total intensity radio image of 3C 397. The image was
deconvolved with the maximum intensity method. Barry Clark CLEANed image had no 
significant difference. The synthesized beam is 6.4 x 5.6\arcsec. The rms noise 
is 1.2 mJy beam$^{-1}$. Primary beam corrected. The cross indicates the location 
of the X-ray hot spot $\alpha$ 19$^h$ 7$^m$ 35\fs13, $\delta$ 
7\arcdeg~8\arcmin~28\farcs76.
\label{cvtc}}
\end{figure}

\placefigure{lvtc}
\placefigure{cvtc}

The total-intensity images presented here have about twice the
resolution of earlier observations. The synthesized beams and
sensitivities of the images are given in Table \ref{summary}.
On the shortest baselines we detected 18 Jy at 20 cm (this includes
both the SNR and the HII region, approximately 4 Jy) and 5.5 Jy at 6
cm.  At 20 cm we should be sensitive to most of the flux from the
remnant, though our value of about 14 Jy is somewhat lower than most
published values (see Fig.~\ref{ford}).
As we state in the Introduction, the huge dispersion in published
fluxes indicates to us that our value of about 14 Jy is not obviously
contradicted by other work.  However, we do suspect that substantial
flux is missing at 6 cm.  In our final cleaned maps we detect, in the
SNR, 5.8 Jy at 6 cm and 13.8 Jy at 20 cm.  A negative bowl does appear
around the remnant in our 6 cm images, indicating missing flux.  The
depth of the bowl, of order 1--2 mJy, would give a total flux deficit
of about 1--2 Jy if distributed uniformly.

As mentioned above, we looked for artifacts due to the deconvolution
by comparing these cleaned images with images deconvolved using the
AIPS maximum-entropy algorithm VTESS.  This routine constructs
maximally smooth maps constrained to fit the data to within a given
noise tolerance.  If the total flux in the source is known, it can be
given as an additional constraint.  In our case, since the total flux
of 3C 397 is very poorly known and the H II region is also in the
field, we allowed VTESS to optimize the fit by varying the total flux.
It returned values of 23.7 Jy at 20 cm and 9.75 Jy at 6 cm, part of
which belongs to the H II region and the positive bias background.
After primary-beam correction, the flux in the remnant measured from
these images was found to be 14.1 and 6.0 Jy at 20 and 6 cm,
respectively.  These values are comparable to those obtained from the
CLEAN maps.  The presence of the large H II region, partially resolved
out at 20 cm and both partially resolved out and suffering from
primary-beam attenuation at 6 cm, made image reconstructions
unreliable at the level of a few mJy beam$^{-1}$.  This is a small amount
compared to the brightness of significant features, but can cause
serious offsets in inferred values of spectral index, as we discuss
below.  We quantified the level of uncertainty due to deconvolution by
subtracting VTESS from CLEAN maps at each frequency.  These maps,
which should be zero, had rms fluctuations of 0.6 mJy beam$^{-1}$ (at 20 cm)
and 0.7 (at 6 cm).  To study the effects of the H II region, we
created and CLEANed a smaller region of sky, including the remnant but
cropping out the faint outer parts of the H II region, at 20 cm.  The
difference between this image and the larger image, both deconvolved
with CLEAN, had an rms of 1.1 mJy beam$^{-1}$.  

\placetable{summary} \placetable{noise} Figure \ref{lvtc} shows the
total-intensity image at 20 cm, with a resolution of about 6\arcsec, 
deconvolved using VTESS (the maximum-entropy method).  The lowest
contour is at 1.8 mJy beam$^{-1}$, 3 times the rms noise.  At 20 cm
the remnant is shaped like a irregular rectangle with enhanced
emission evident along the southwest rim, as well as at locations
interior to the southwest rim and within the southeast emission. The
image agrees in overall structure with 14\arcsec~resolution images
observed by \cite{Becker1985} (1985) and \cite{AR1993} (1993). Part of
the southern edge may be unresolved, while some emission can be
detected outside of the rim at the southwest corner of the
remnant. The 6 cm image (Figure \ref{cvtc}) is nearly identical to the
20 cm image in structure, implying little variation in spectral index
within the remnant.  Little structure appears to be still unresolved
at this resolution, with the possible exception of the emission at
some portions of the remnant edge.

Figure \ref{cvtc}, the 6-cm image deconvolved with VTESS, is almost
indistinguishable from the 20-cm image, 
Figure \ref{lvtc}.  
A scaled subtraction of the images (the 20-cm image less 2.2 times the
6-cm image) minimized the rms difference at 1.3 mJy beam$^{-1}$, the level
of uncertainty in flux baselines due to missing interferometer 
spacings.  The factor of 2.2 over a frequency range of a factor of 3.3
gives a mean spectral index of --0.66.

\begin{figure}
\plotone{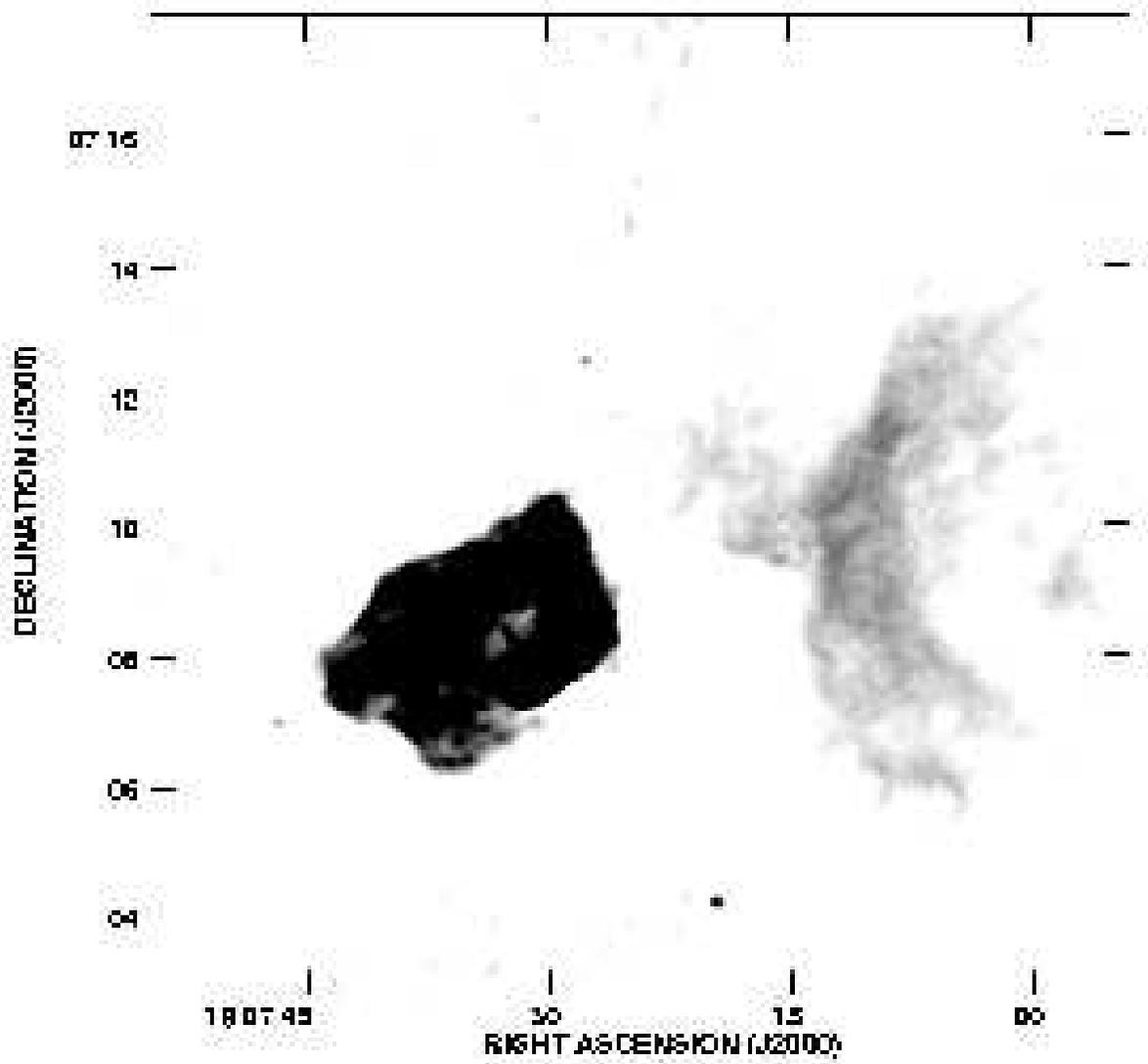}
\caption{20 cm radio image showing HII region to the northwest of 3C 397.
The beamsize is 6.4 x 6.1\arcsec. \label{HII}}    
\end{figure}

Figure \ref{HII}, a 20-cm image, shows the H II region which lies to
the northwest of the remnant. We detect a flux of 4.3 Jy at 20 cm from
this region.  Since it is too large to be fully resolved, this is a
lower limit.  As described above, H I absorption evidence suggests
that the H II region is closer than the remnant, and our images do not
show any morphological evidence of interaction.  While we cannot rule
out an association, we find it unlikely and shall assume the H II
region is a foreground object.

If the observed H II region has an extended relatively low-density
envelope, it could cover 3C 397 and account for the observed
low-frequency absorption without contributing observable thermal
emission.  Even at the far distance of 9.3 kpc, an envelope radius of
30 pc from the brightest part of the observed H II region would cover
3C 397.  A free-free optical depth of 1 at 30 MHz, as observed (Kassim
1989b), could be obtained with a temperature of 5000 K, a
line-of-sight dimension of 30 pc, and a number density of 5 cm$^{-3}$.
These parameters would imply an emitted intensity of only 3 $\mu$Jy
arcsec$^{-2}$ or about 0.1 mJy per $6^{\prime\prime}$ beam, well below
our noise levels.

	\subsubsection{Polarization}

Images of Stokes parameters $Q$, $U$ and $V$ were made at
both 6 and 20 cm to check for linear polarization (See \cite{II} \S 3.2 for
a discussion of circular polarization in SNRs). The $Q$ and $U$ images
were convolved to a resolution of 15\arcsec~to increase
sensitivity to extended emission. Polarized-flux images were created
following standard procedures. A total polarized flux of 0.15 Jy was
detected at 6 cm. Using a total flux density of 10.3 Jy at 6 cm from
\cite{Green1998}'s mean spectrum, we found the mean polarized fraction
(i.e. the total polarized flux divided by total flux) measured from 3C 397
at 6 cm to be 1.5\% $\pm$ 0.1\%.  Alternatively, if we use our
measured total flux of 5.6 Jy, we obtain 2.7\% $\pm$ 0.1\%.  These values
should bound the true polarized fraction.

\begin{figure}
\plotone{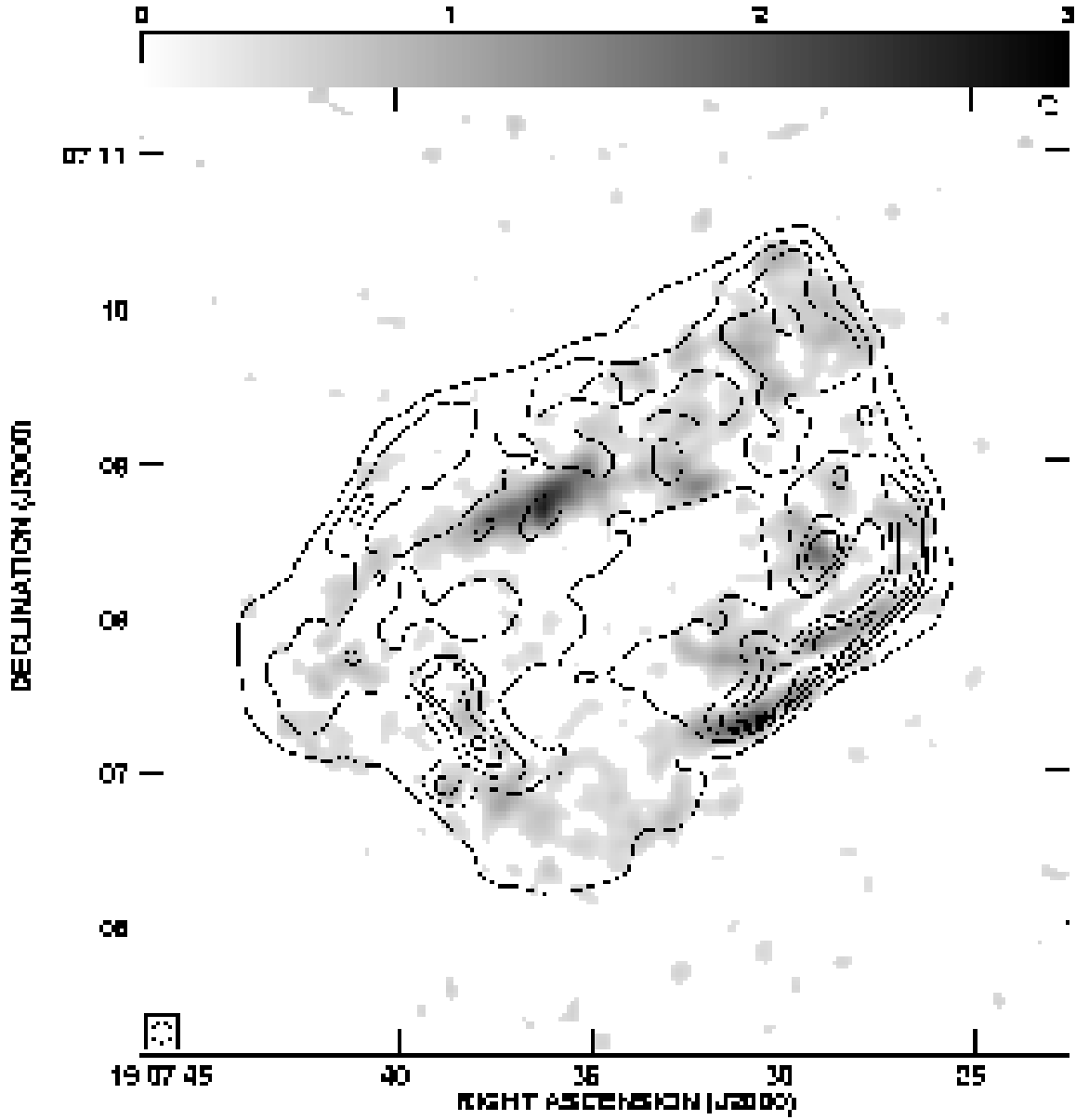}
\caption{6 cm polarization intensity of 3C 397 shown in grayscale (0.4 to 2 mJybeam$^{-1}$) with overlaid 6 cm total intensity contours. Both images are convolved to 8\arcsec. Contours are at the level of 5, 15, 25, 34, 45, 55$\sigma$
where $\sigma=$0.61 mJy beam$^{-1}$. \label{cpol}}    
\end{figure}
\begin{figure}
\plotone{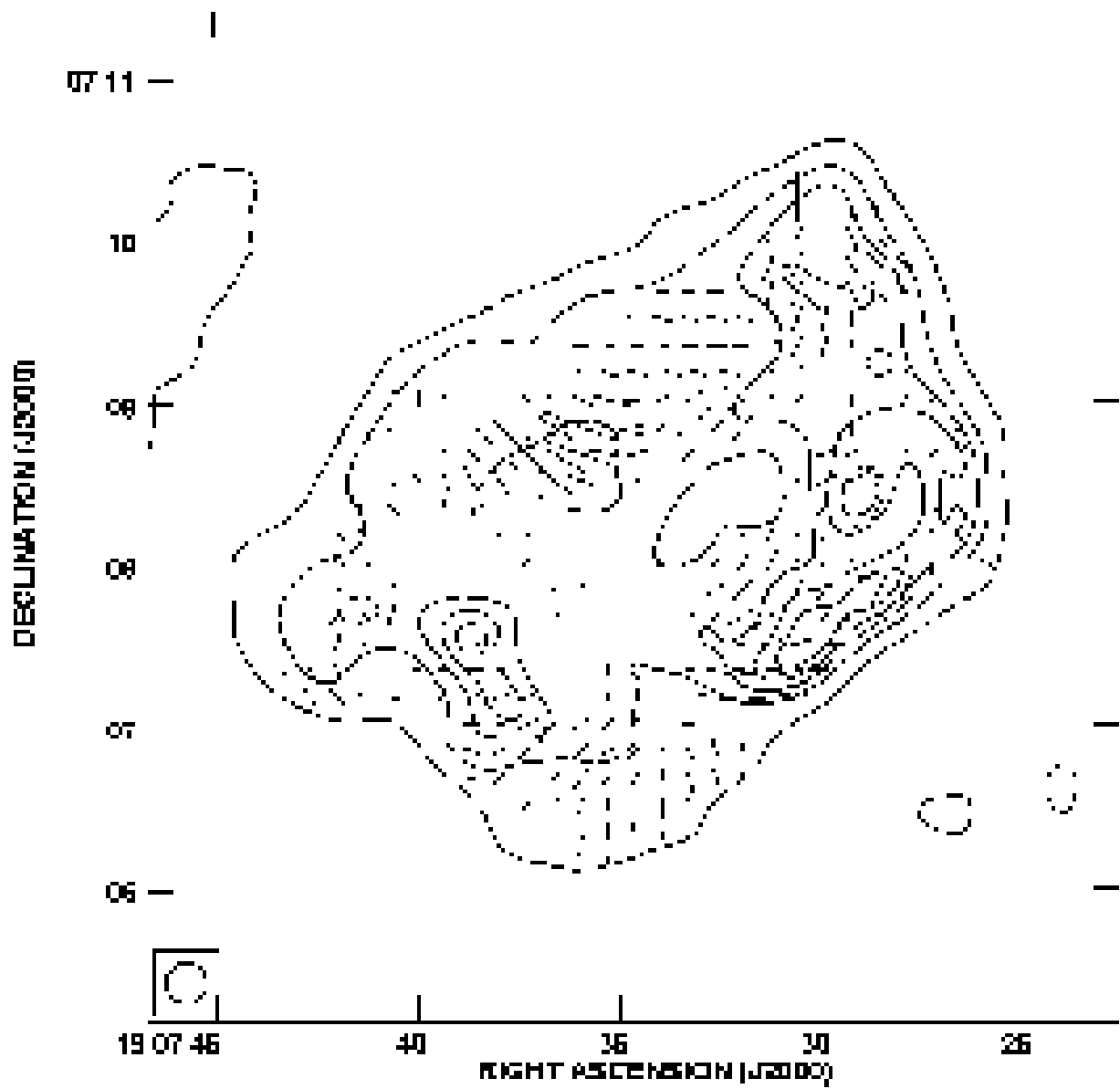}
\caption{6 cm polarization vectors representing the measured electric
field in 3C 397. The length of the vectors represents the strength of the
polarization and the relative orientation the angle. Angles have not been corrected
for Faraday rotation. Contours are at 5, 20, 35, 50, 65, 
80$\sigma$ where $\sigma$= 0.13 mJy beam$^{-1}$.  All images are convolved to 15\arcsec. \label{pola}}    
\end{figure}
\begin{figure}
\plotone{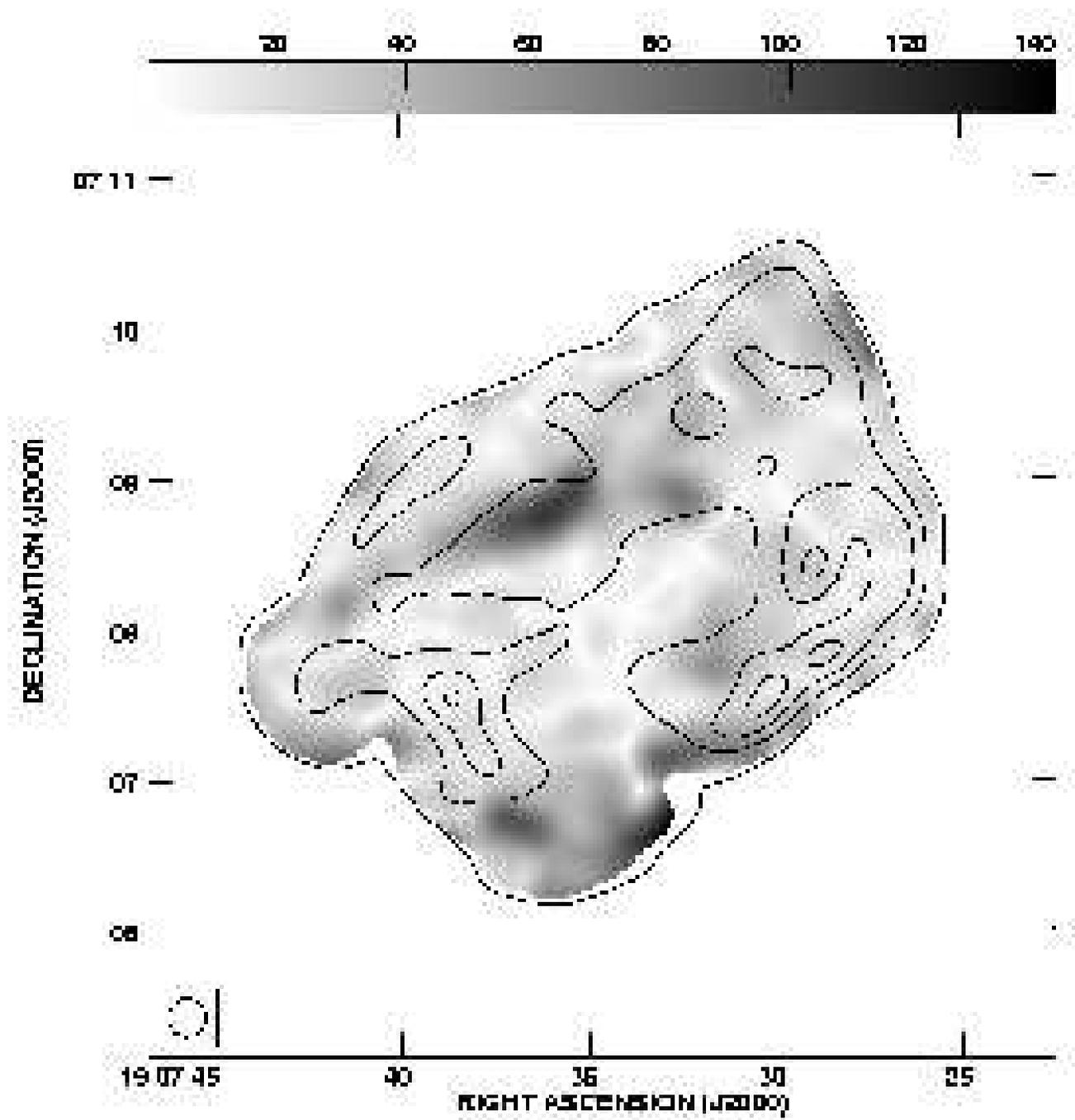}
\caption{Fractional polarization image of 3C 397 at 6 cm. Grayscale is 
fractional
polarization in units of millipercent, contours are 6 cm total intensity
at 5, 15, 25 and 36 $\sigma$. The beamsize for both is 
15\arcsec.\label{frac.pol}}    
\end{figure}

No polarization was detected at 20 cm to a limit of $1.2 \times 10^{-4}$
Jy beam$^{-1}$ (corresponding to $3\sigma$). 
Due to the lack of detection at 20 cm we cannot remove
Faraday rotation from the 6 cm observed electric-vector position
angles.  Figure \ref{cpol} shows a grayscale image of the linearly
polarized intensity at 6 cm, superimposed on a simple contour map of the total
intensity.  Figure \ref{pola} shows the polarized intensity as
polarization vectors, overlaid on a total-intensity contour map, both
convolved to 15\arcsec. Figure \ref{frac.pol} shows a grayscale
image of the fractional polarization, overlaid on a total-intensity
contour map, convolved to 15\arcsec.  Polarization structure
does not correspond well to structure in total intensity.  Peak
polarized-fraction values approach 11\%, excluding regions near
the remnant edge likely due to noise.

	\subsubsection{Spectral Index} 
\begin{figure}
\plotone{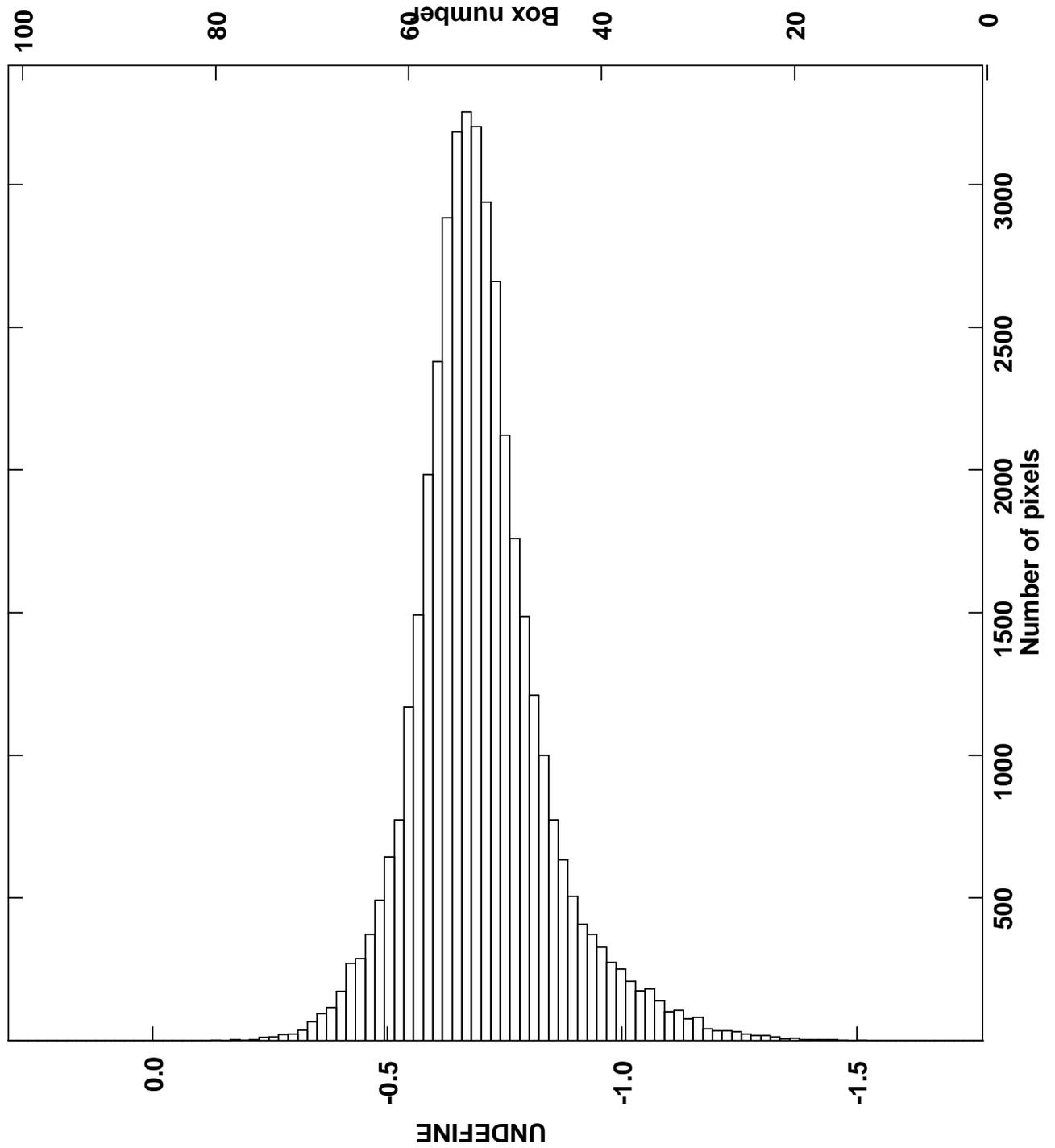}
\caption{A  histogram of all values of spectral index for 3C 397 from regions of 
the CLEANED map brighter than 2.2 mJy. Convolution for the spectral index map 
was 6\arcsec. \label{spixhist}}
\end{figure}

\begin{figure}
\plottwo{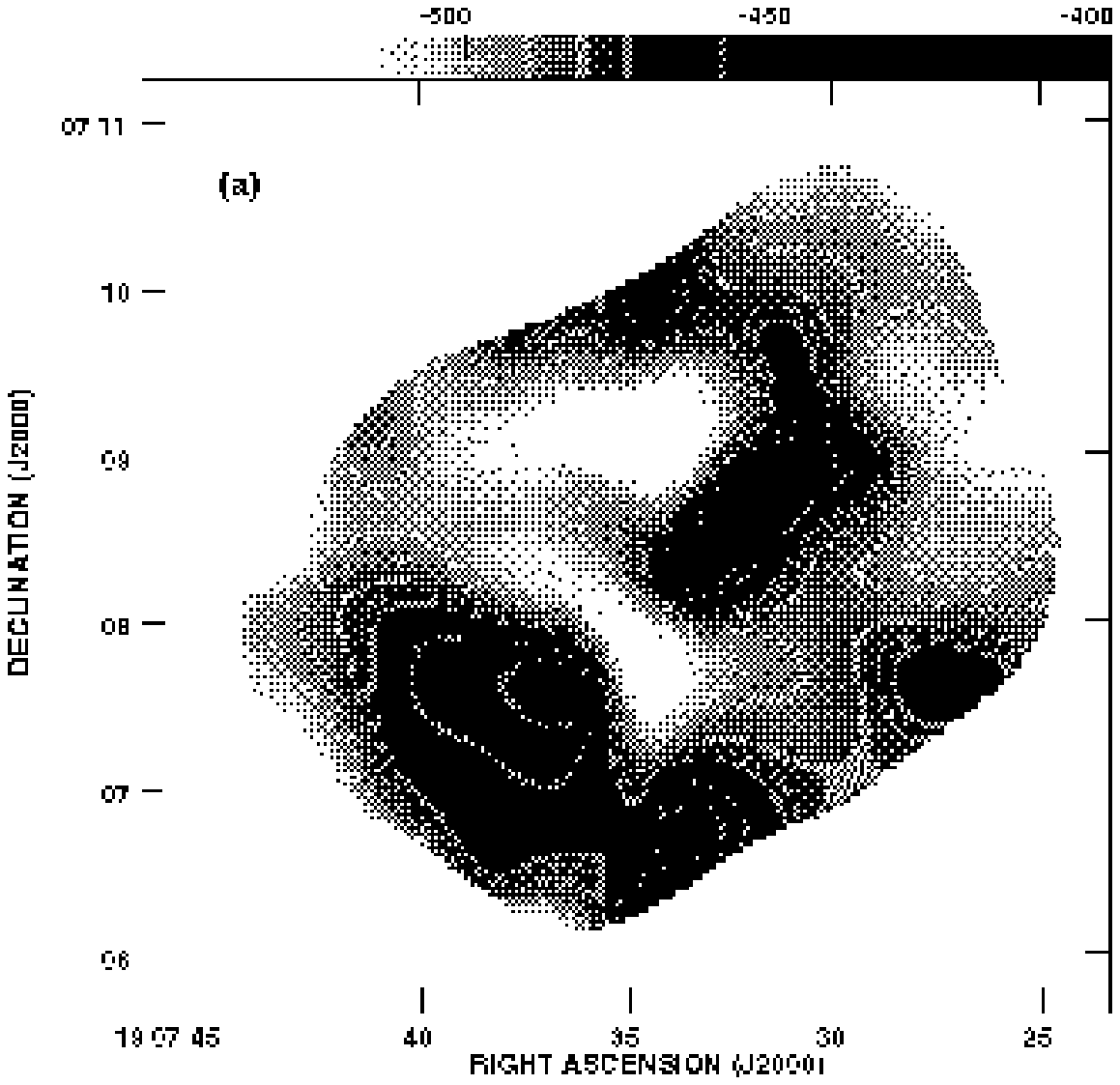}{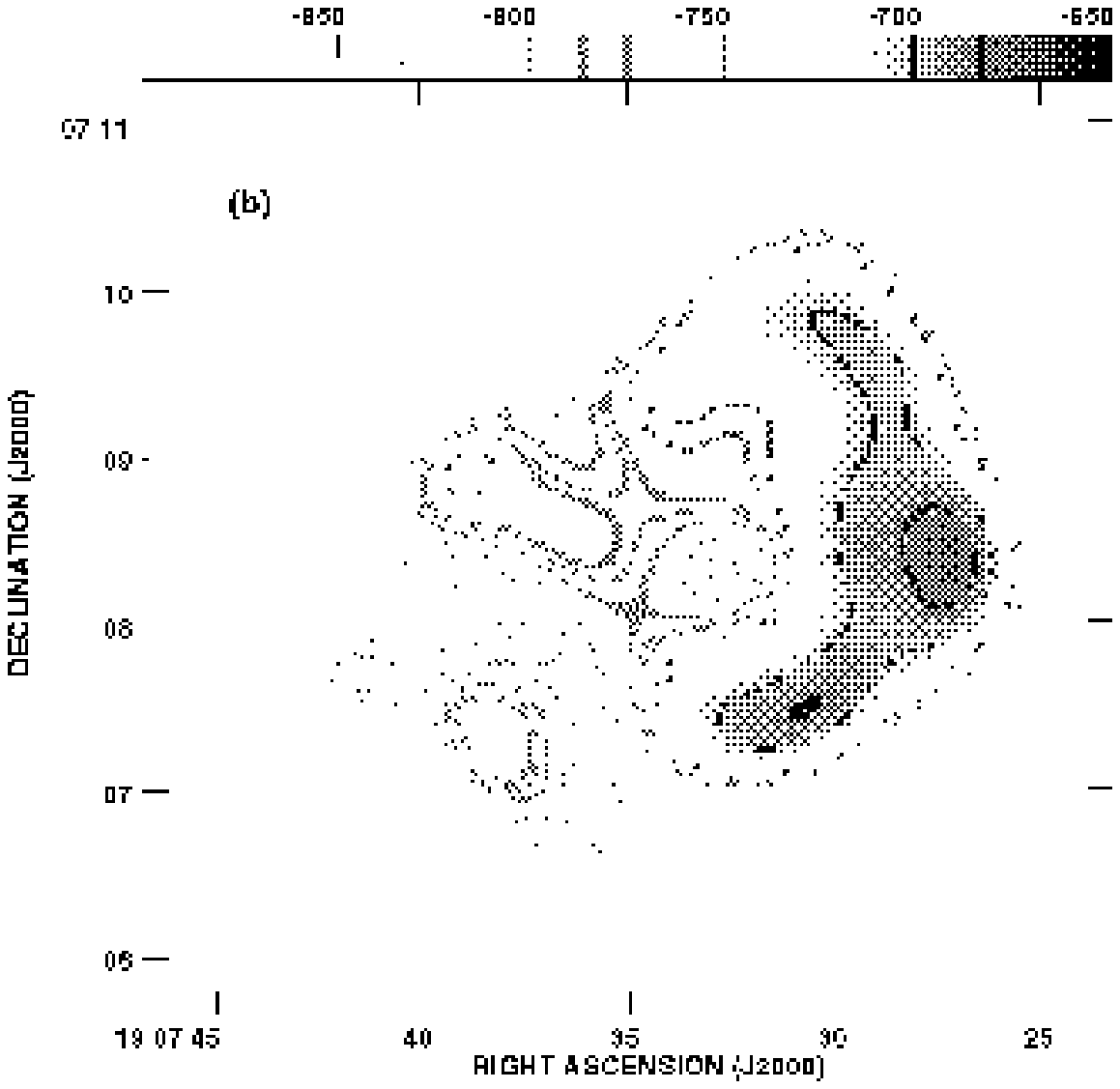}
\caption{(a) Spectral index image of 3C 397 between 6 cm and 20 cm,
from the regression method (see text) with effective resolution
of 40\arcsec. (b) Spectral index image from $\log(S_2/S_1)/\log(\nu_2/\nu_1)$,
at the same resolution.
 \label{MX.M}}    
\end{figure}

\placefigure{MX.M}
\placefigure{spixhist}

Given the large uncertainties in total flux of the remnant at either
frequency, calculating spectral index was problematic.  The integrated
fluxes we observed with CLEAN (13.8 and 5.6 Jy at 20 and 6 cm,
respectively) imply a spectral index for the whole remnant of
$\alpha=-0.76$.  However, the VTESS fluxes of 14.1 and 6.0 give
--0.72.  These are considerably steeper than the spectral index quoted
in \cite{Green1998}'s catalog of $\alpha=-0.48$ from single-dish
measurements, or the value of --0.4 given by \cite{Kassim1989B} (1989b) (though both these values are partly based on  
confused data).  Missing a uniformly distributed component of flux at
the higher frequency produce both an excessively steep overall
spectral index, and an artificial correlation of steeper spectra with
fainter regions.  The scaled subtraction of a multiple of the 6-cm
VTESS map from the 20-cm map produced a difference image whose rms was
minimized for a reduction factor of 2.2, implying a mean spectral
index of --0.66 for the smaller-scale features.  The same procedure
with the CLEAN images gave --0.62.  This method is impervious to
constant offsets at either frequency, in the absence of strong real
spectral-index variations.  The result supports the idea that the true
spectral index is considerably flatter than the values we obtain from
our total-flux determinations.  We are, however, unlikely to be
missing more than a few Jy at 6 cm.  If the true spectral index were
--0.5 and our 20-cm flux of 14.1 Jy correct, we would be missing about
1.8 Jy at 6 cm, corresponding to only 1.5 mJy per 8\arcsec \ beam.
This value is less than the rms of the optimized scaled-subtraction
difference map, and only slightly larger than the typical variations
due to deconvolution (see above).  In summary, we cannot determine a
mean spectral index reliably from our data, but a value of roughly
-0.6 would be consistent with our results.  The flatter values of 
Green and Kassim may be due to use of some confused data. 

\begin{figure}
\plotone{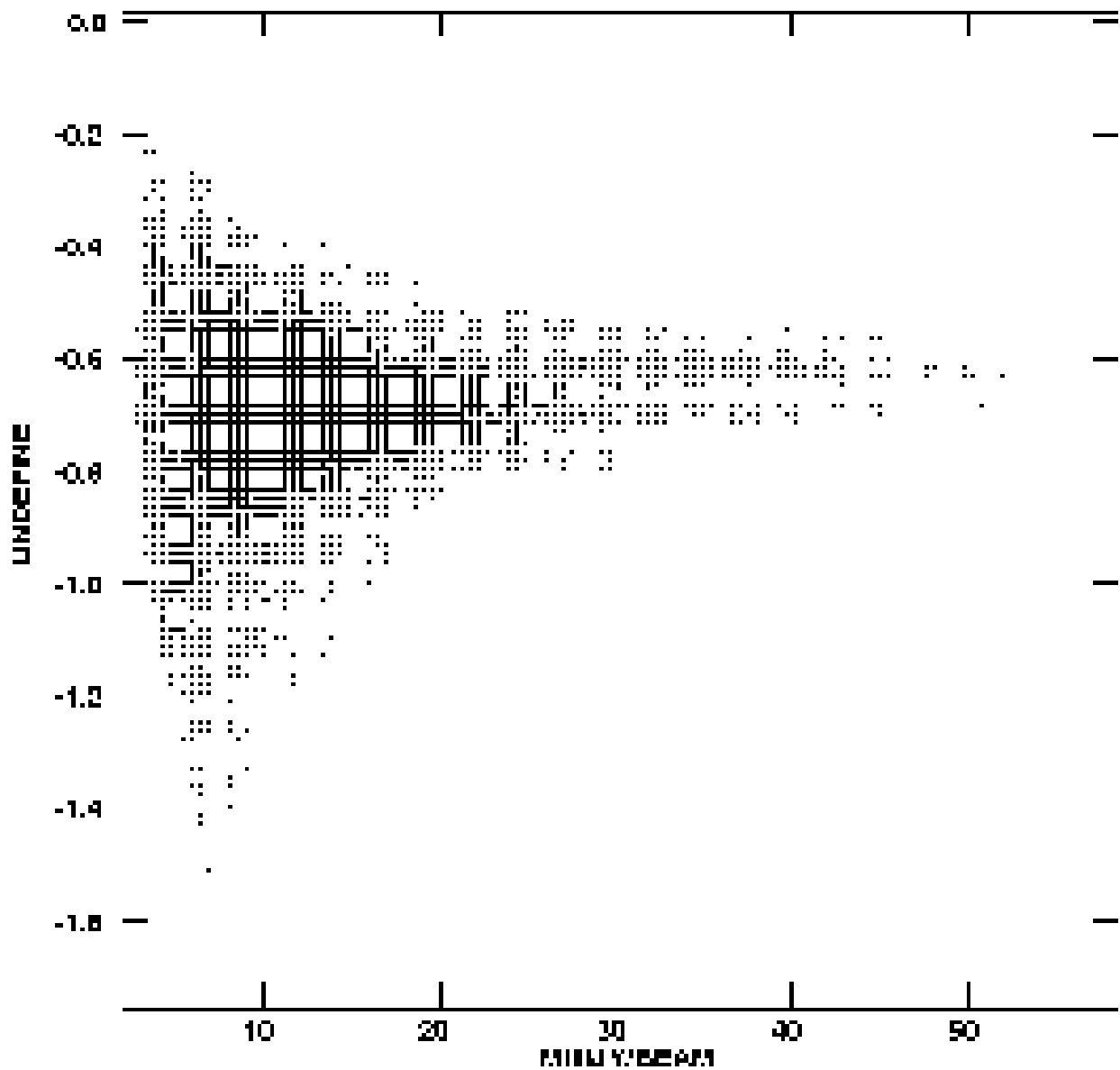}
\caption{Scatter plot of spectral index $\alpha$ vs. intensity for 3C 397. Data
convolved to 6\arcsec.\label{imvim}}    
\end{figure}

We initially created a standard spectral-index map (Figure \ref{MX.M})
whose pixels are $\log (S_6/S_{20}) / \log(\nu_2/\nu_1)$, even though
this formulation amplifies noise due to faint regions.  We failed to
find convincing variations of spectral index associated with any
structures in radio or X-ray emission.  No obvious trends of spectral
index with brightness surfaced statistically, either.  Figure
\ref{imvim} shows a scatter plot of pixel spectral index
vs.~brightness at 20 cm.  The dispersion is substantial, especially at
low flux levels, but above about 5 mJy beam$^{-1}$, the mean of the
distribution appears not to vary appreciably in spectral index as
brightness increases.  Figure \ref{spixhist} shows a histogram of all
values of spectral index from regions of the 20 cm map brighter than
2.2 mJy beam$^{-1}$.  The distribution appears smooth and consistent
with normally distributed errors, even though the distribution's FWHM
of about $\Delta \alpha \sim 0.2$ is considerably larger than the
expected uncertainties of order 0.04 derived from our observed 20 cm
and 6 cm flux uncertainties.  There is no obvious suggestion of
bimodality in the distribution as we might expect if several
physically distinct regions of different spectral index were present.

However, we also examined the data using an alternative method, that
of \cite{AR1993} \ (1993), in which the spectral indices
of regions are found by fitting regression lines to the plots of pixel
brightnesses at 20 cm vs.~at 6 cm over regions of specified size.
We used a region corresponding to a Gaussian-weighted
average with a half-width of 40\arcsec.  The method finds the slope
and intercept of the best straight-line fit.  Errors can be estimated
by comparing the values for the regression of 20 on 6-cm values with
the reverse regression of 6 on 20-cm values.  Since fractional errors
are larger on the 6-cm image (typical fluxes less by a factor of 2.2,
but noise less by only 1.2 to 1.5), one expects an asymmetry.  This
method was applied both to CLEAN and VTESS images; mean values of the
difference images between the two regressions were 0.05 and 0.04 in
spectral index, respectively, with rms scatter of 0.04 and 0.03,
respectively.  We conclude that an intrinsic dispersion of $\pm 0.05$
unavoidably characterizes the spectral-index determinations.  The
spectral-index images from CLEAN and VTESS maps differ by less than
the two regressions on one set of data (rms 0.033).  Figure \ref{MX.M}
shows only minor variations; the steepest regions, at the edges, are
almost certainly artifacts due to extra missing flux at 6 cm.  Figure
\ref{MX.M} also shows a traditional spectral-index image made from
CLEANed images convolved to a resolution of 40\arcsec.  There
appears to be no correlation between the two spectral-index images.
We conclude that our ability to discriminate true spectral-index
variations is at about the $\pm 0.05$ level, and that to this level,
we do not see obvious variations across 3C 397.


        \subsection{X-ray}

\begin{figure}
\plotone{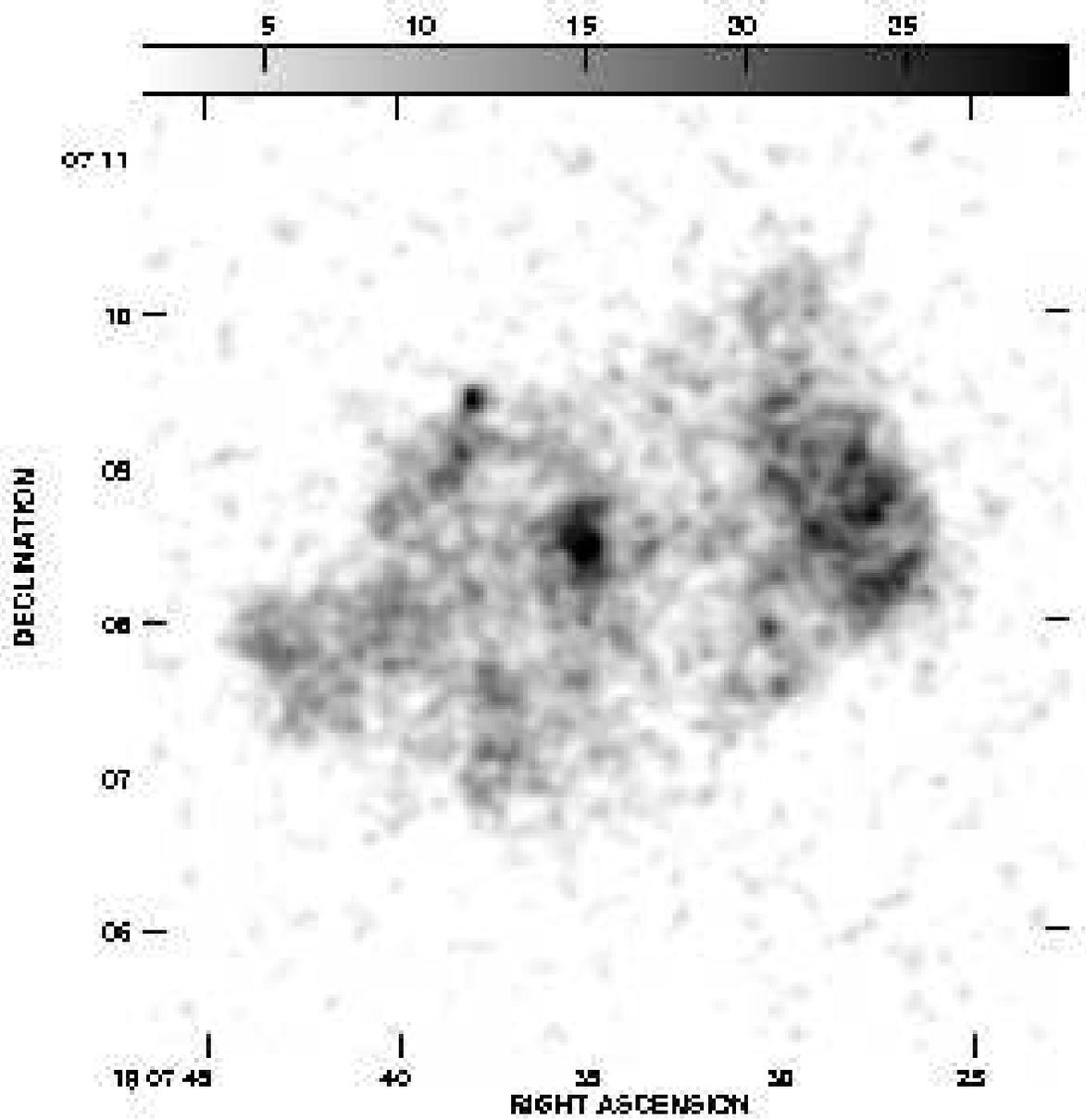}
\caption{ROSAT {\it HRI} image of 3C 397. Convolved to 6\arcsec~with background 
subtracted. Units are in counts/beam.  \label{HRI}}

\end{figure}

\begin{figure}
\plotone{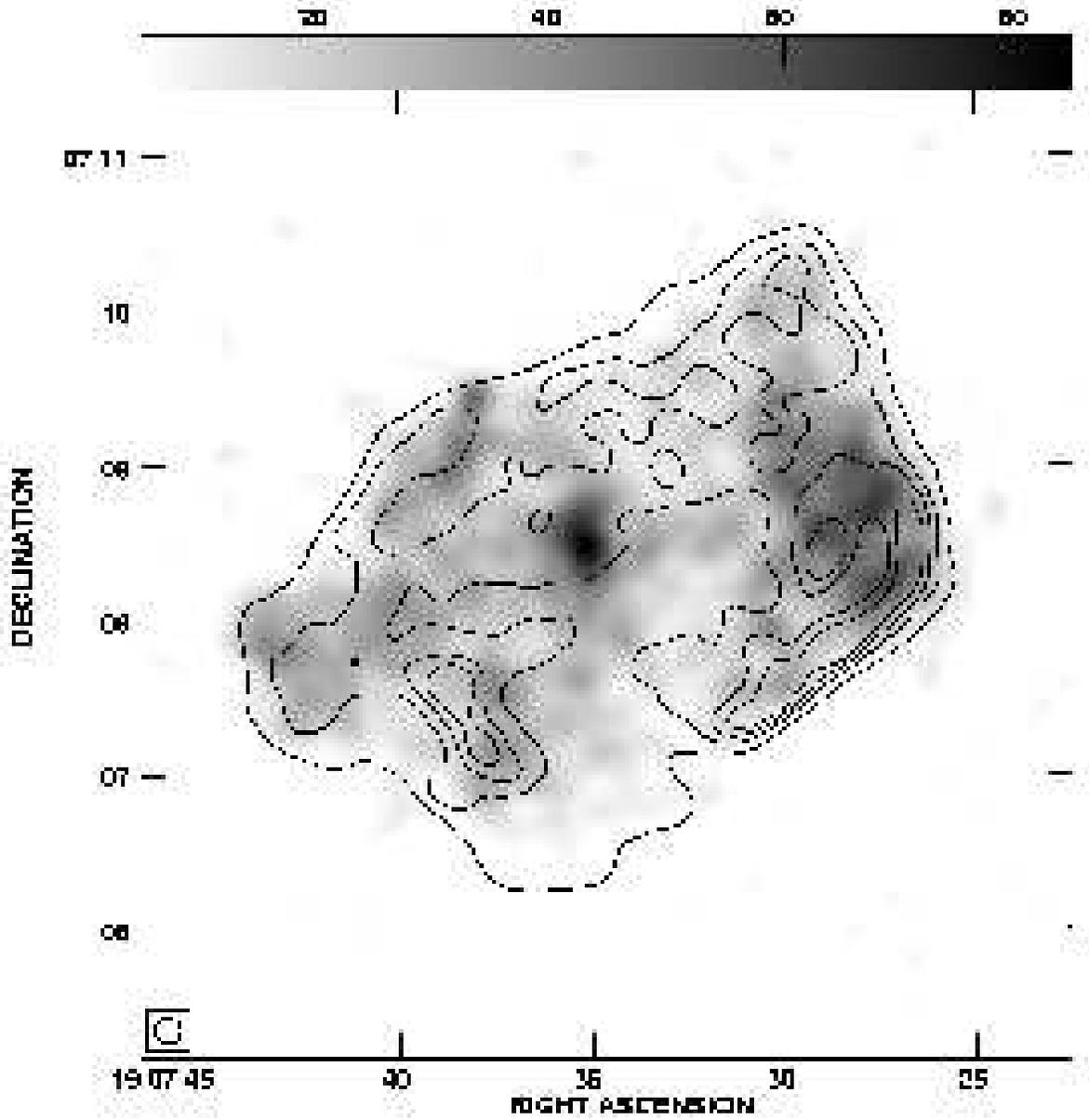}
\caption{Grayscale of ROSAT {\it HRI} X-ray emission, smoothed to 10\arcsec.
Contours are VLA Observations of 3C 397 at 20 cm (beam 10\arcsec). The
contours are 15, 45, 75, 105$\sigma$ where $\sigma$ = 0.71 mJy. Note the X-ray 
hot spot, which does not correspond to any emission structure in the 
radio.\label{L+X}}    
\end{figure}

The most striking feature of the {\it HRI} image (Figure \ref{HRI}) is the
strong resemblance between radio and X-ray morphologies.
Figure \ref{L+X} shows radio contours superposed on X-ray grayscale,
both smoothed to 10\arcsec~resolution.  The X-ray and radio edges
appear coincident over most of the remnant, though in some regions
such as the southern extremity the X-ray emission is relatively
fainter than radio and appears to fade into the noise inside the
radio edge.  The brightest radio region is in the southwest corner,
where strong X-ray emission is also concentrated. 

\begin{figure}
\plotone{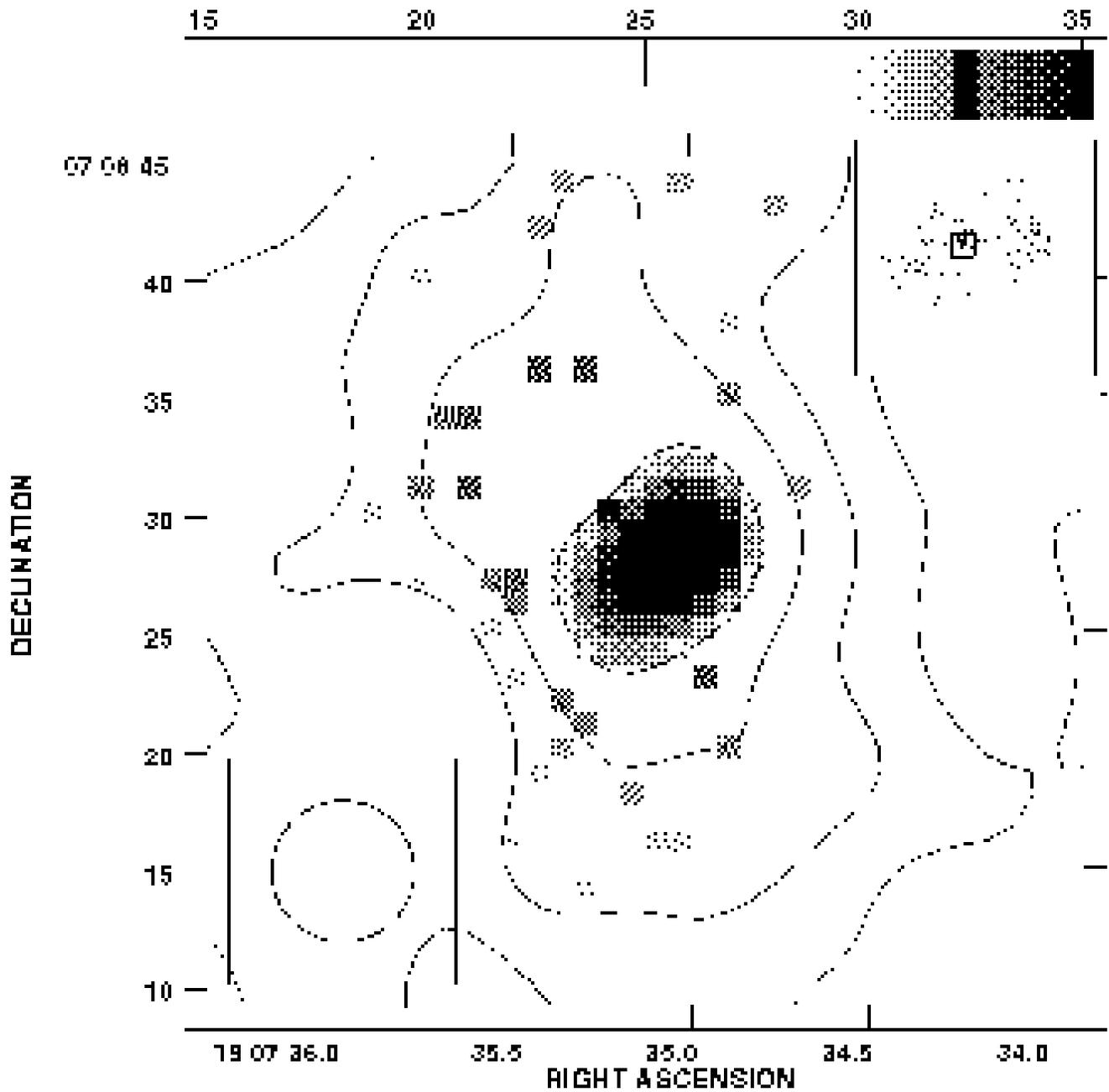}
\caption{The X-ray hot spot in 3C 397. Grayscale ROSAT {\it PSPC} convolved to 
6\arcsec~ with contours at 10, 15, 20, 25$\sigma$ where $\sigma$ = 1.2 
counts/convolved beam. \label{HOTSPOT}}
\end{figure}

The most obvious difference
between the X-ray and radio images is a bright compact central feature in
the X-ray image, which has no counterpart in the radio. A Gaussian fit to the 
X-ray hot spot places it at $\alpha$ 19$^h$ 7$^m$ 35\fs13, $\delta$ 
7\arcdeg~8\arcmin~28\farcs76. This feature, discussed more fully in 
\S~\ref{sec:Discussion}, is shown in detail in Figure \ref{HOTSPOT}.
Its flux was estimated assuming the thermal
spectrum described below:  a poor fit, but adequate for an estimate.
We obtain a flux between 0.4 and 2 keV, corrected for absorption, 
of about $3 \times 10^{-12}$
erg s$^{-1}$ cm$^{-2}$, or 3--5\% of the remnant total (see below).
At 10 kpc this is about $4 L_\odot$.  
The feature appears to consist of an unresolved core and resolved,
roughly north-south extensions in both directions.  The extensions
span about 30\arcsec.

We used the ROSAT {\it HRI}'s millisecond time resolution to search for
pulsations from this bright region. Selecting photons from the region
a circular region of radius 19\arcsec~about $\alpha$ 19$^h$ 7$^m$ 34\fs86, 
$\delta$ 7\arcdeg~8\arcmin~31\farcs0~we searched for periodicity, using 
coherent fast Fourier transforms,
after applying the standard barycenter corrections.  The results were
consistent with a non-periodic signal.



\begin{figure}
\plotone{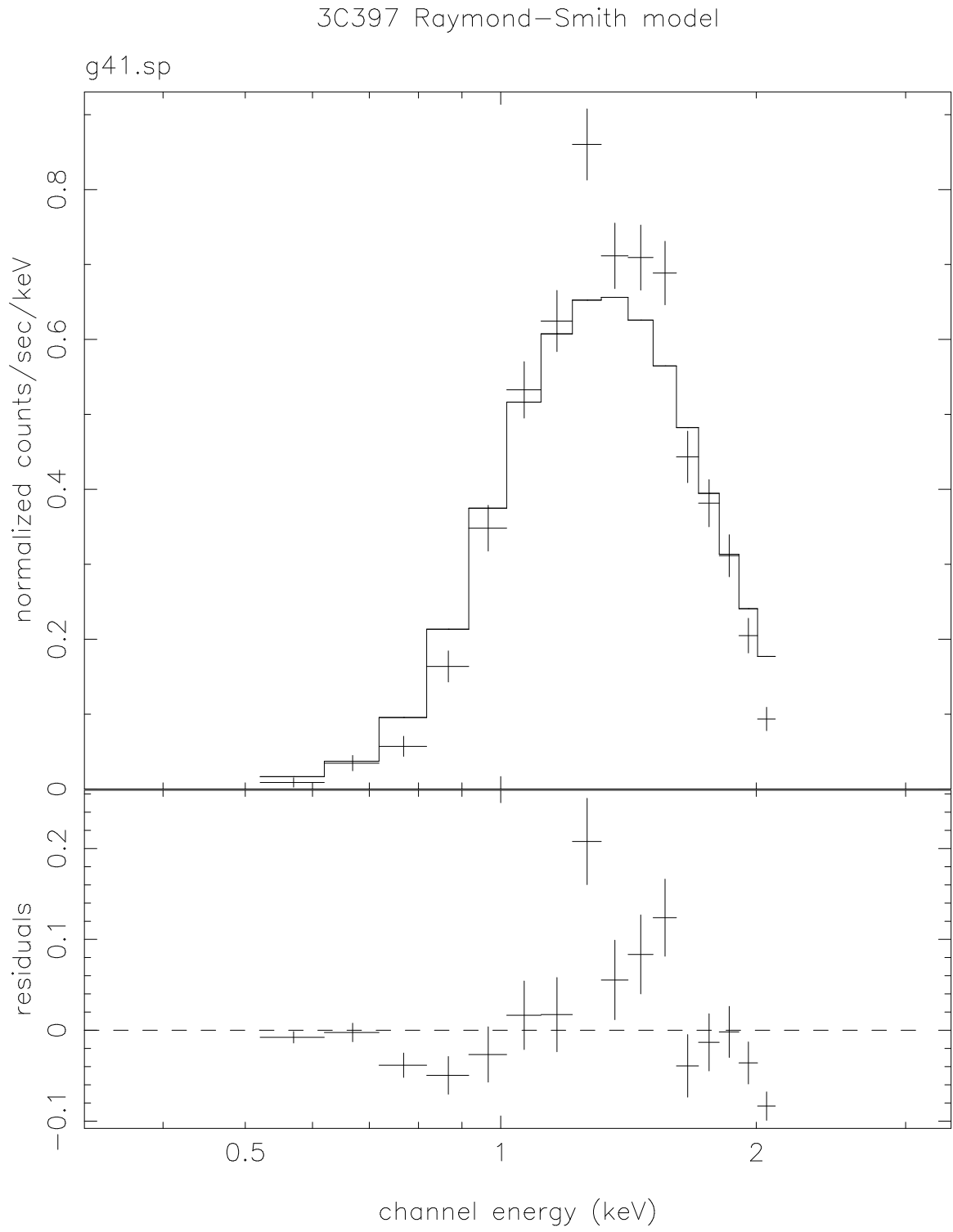}
\caption{3C 397 ROSAT {\it PSPC} spectrum. \label{xspec}}
\end{figure}

The {\it PSPC} spectrum is shown in Figure \ref{xspec}.  Strong absorption is
obvious, as should be expected for a distance of order 10 kpc (which results in
a Galactocentric distance of 6.6 kpc, for a Solar Galactocentric radius
of 8.5 kpc).  

The data do not justify more than the simplest spectral
fits; a Raymond-Smith equilibrium plasma model, which we expect to be
a poor description for a relatively young remnant like 3C 397, yields
a temperature of 1.8 keV and an absorption column of $1.3 \times
10^{22}$ cm$^{-2}$, with a totally unacceptable reduced $\chi^2$ of
6.3.  Similar poor results are obtained for other plausible models: a
one-component homogeneous nonequilibrium-ionization model (plasma at a
constant temperature and density for a time $t$) does even worse, and
a plane-shock model (a superposition of components of different ages
and temperatures to simulate post-shock emission; Borkowski, in
preparation) is worse yet.  Oddly, a good description of the spectrum
is provided by a single power-law model with a column $N_H = 3.2
\times 10^{22}$ cm$^{-2}$, and an absurd energy spectral index
$\alpha_x$ of --7.6 ($S_x \propto \nu^{\alpha_x}$).  We have thus not
invested any effort in attempting to extract physically meaningful
results from the {\it PSPC} spectrum, especially since ASCA spectra are
available.  A more complete discussion of the X-ray spectrum of 3C 397
will have to wait for more sophisticated models in conjunction with
those data.

Our spectral results are similar but not identical to those of \cite{Rho1998} 
(1998), who used the same dataset; small differences in binning
and background subtraction account for the differences.  We were also
unwilling to invest effort in adjusting abundances, etc., given that
the much better ASCA data are available.

\begin{figure}
\plotone{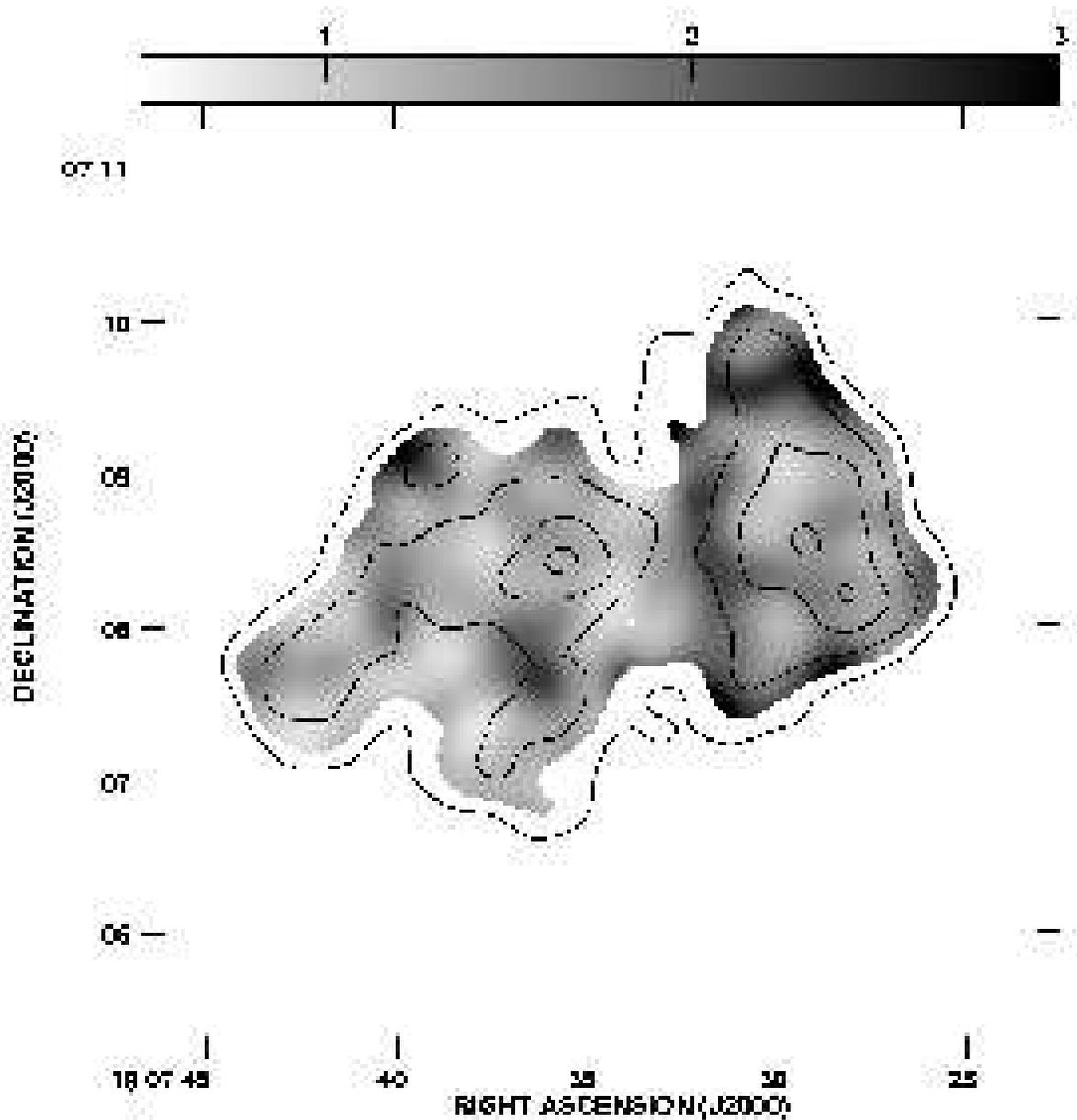}
\caption{Grayscale ROSAT {\it PSPC} hardness-ratio map of 3C 397 with X-ray brightness contours. This map
was constructed by taking the ratio of the image of counts above 1.3 keV to the 
image below 1.3 keV. It is smoothed with a $20\arcsec$ Gaussian.  It shows only 
areas where the count rate is more than 3 times the mean background. Contours of 
X-ray brightness are 0.5, 1, 1.5, 2 counts/convolved beam.\label{hardx}}
\end{figure}
\begin{figure}
\plotone{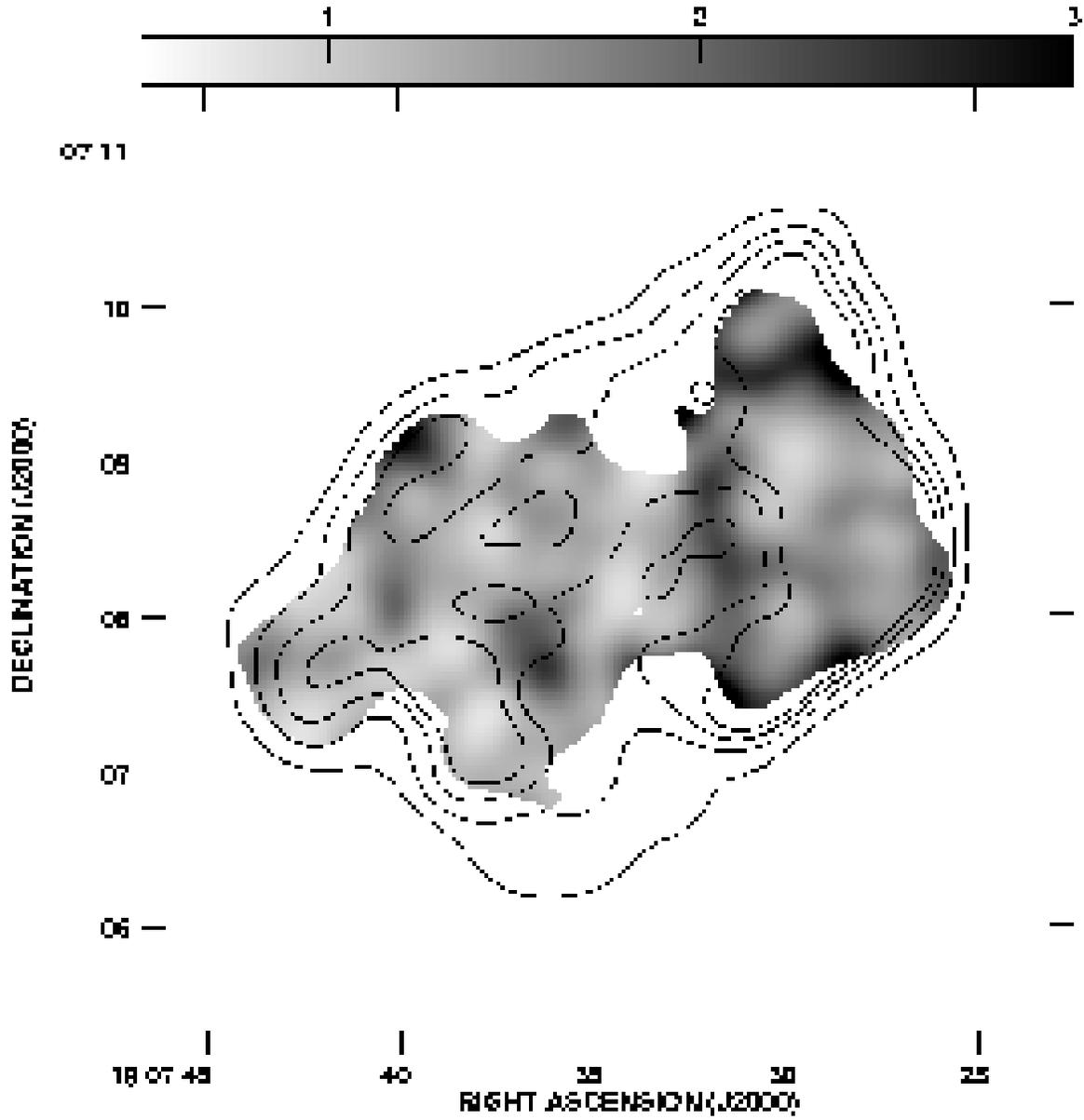}
\caption{Grayscale ROSAT {\it PSPC} hardness-ration map (See Figure \S \ref{hardx}) of 
3C 397 with 20 cm radio contours. Both are convolved to 20\arcsec.  Contours of 
20 cm emission are 20, 50, 80, 110, 140$\sigma$ where $\sigma$ = 1.7 mJy.  
\label{hardr}}
\end{figure}

The {\it PSPC} image is completely consistent with the smoothed {\it HRI} image
and is not shown.  However, we searched for spatial variations of the
X-ray spectrum by constructing a hardness-ratio map, the ratio of
counts above 1.3 keV to those below (Figure \ref{hardx}).  This map
was constructed by taking the ratio of the two appropriate images
smoothed with a $20\arcsec$ Gaussian.  It shows only areas where the
count rate is more than three times the mean background.  The variations
in the map are small and likely to be statistical.  The superposed
contours in Figure \ref{hardx} show that there is no correlation
between hardness and total X-ray brightness, while Figure \ref{hardr}
shows no correlation with radio brightness.  In particular, the
possibly unresolved point source in the {\it HRI} image is not associated
with a harder X-ray spectrum than average.  There were not enough
counts in that region of the {\it PSPC} image alone to attempt spectral
fitting.  We should caution the reader than another remnant of similar
size and brightness, G11.2--0.3, also showed no hardness variations in
the ROSAT data (\cite{Reynolds1994} 1994) while actually harboring a
power-law, pulsar-driven component detectable only in the ASCA data
(\cite{Vasisht1996} 1996).

\section{Discussion}
\label{sec:Discussion}


\subsection{Morphology}

Our X-ray images and spectra have shown conclusively that 3C 397 is
not a thermal composite or mixed-morphology remnant, since there is a
strong, though not perfect, resemblance between the X-ray and radio
images, including substantial edge brightening in X-rays (except for
the X-ray unresolved source, discussed below).  In classic thermal
composites like W49 B (\cite{II} \& \cite{Pye1984} 1984), the X-ray
emission shows no hint of shell structure corresponding to the outer
radio contours (\cite{II}).  If 3C 397 contained both a thermal shell
and a thermal interior of different properties (perhaps a thermal
composite in the making), one might expect spectral gradients in
X-rays, which we do not see in our hardness-ratio maps.  We attribute
the lack of stronger limb-brightening to the same superposition
effects to which we attribute the radio morphology.

Similarly, our observations confirm that 3C 397 does not contain a
plerionic component.  Our spectral-index images rule out a
flat-spectrum component, as do those of earlier studies.  Our
polarimetry shows that the peak of polarization is quite weak and not
coincident with the peak of radio brightness.  The mysterious X-ray
source is coincident with neither.  (At the location of the X-ray
hot spot, the polarized fraction is 0.07.)
 We can thus classify 3C 397 firmly
as a shell remnant with good radio and X-ray correspondence.

\begin{figure}
\plotone{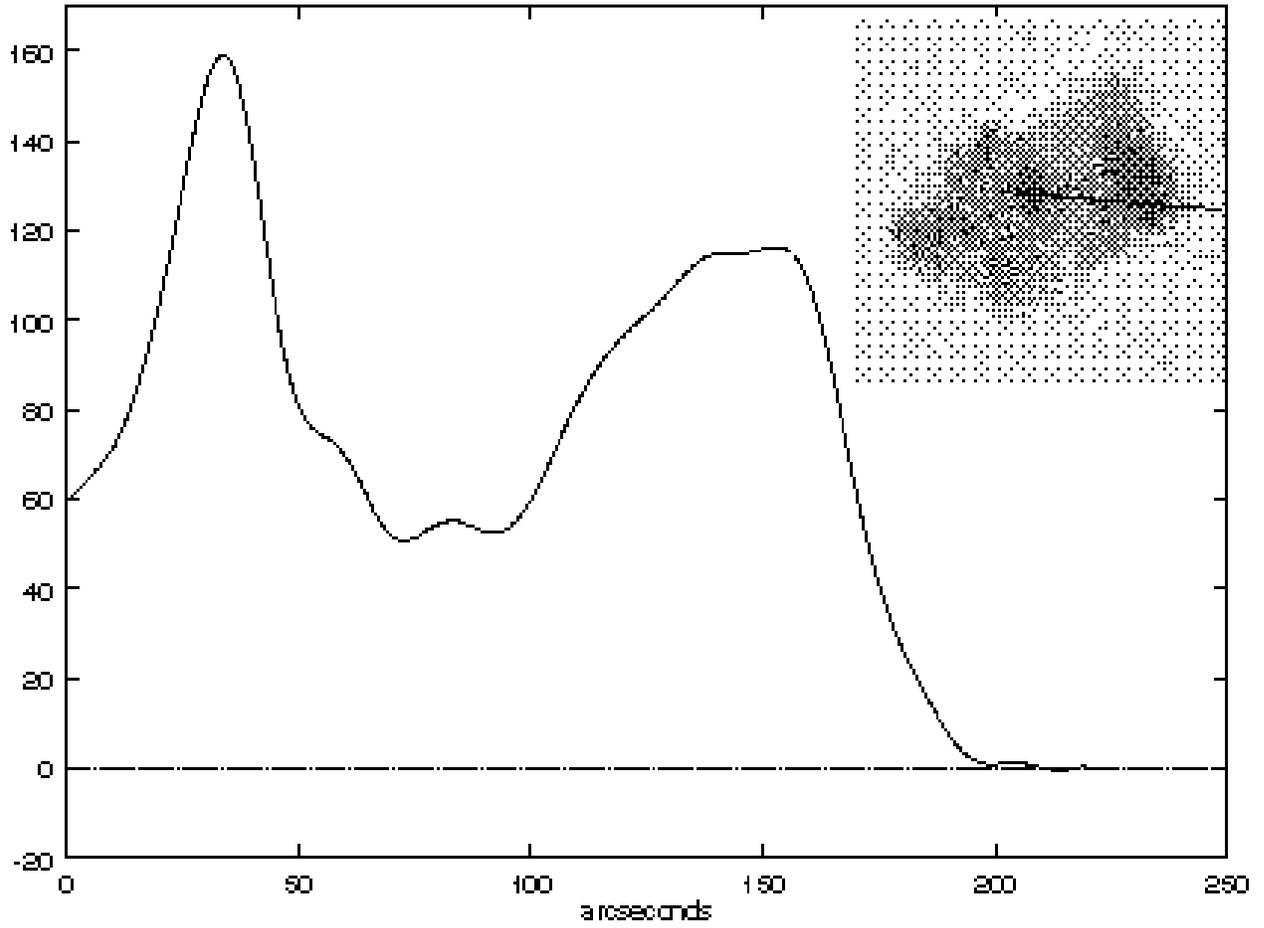}
\caption{Plot of intensity through the hot spot and southwest edge of SNR 3C 
397. \label{slice}}
\end{figure}

\placefigure{slice} 

The unusual non-circular morphology of 3C 397 is most likely a result
of a non-uniform ISM, though there is no direct evidence for molecular
gas in close proximity.  3C 397 is one of the least circular of the
ten Galactic SNRs with highest radio surface brightness.  The
coincidence of radio and X-ray edges supports the idea that the radio
morphology reflects the general hydrodynamics rather than some kind of
synchrotron pathology.

\begin{figure}
\plotone{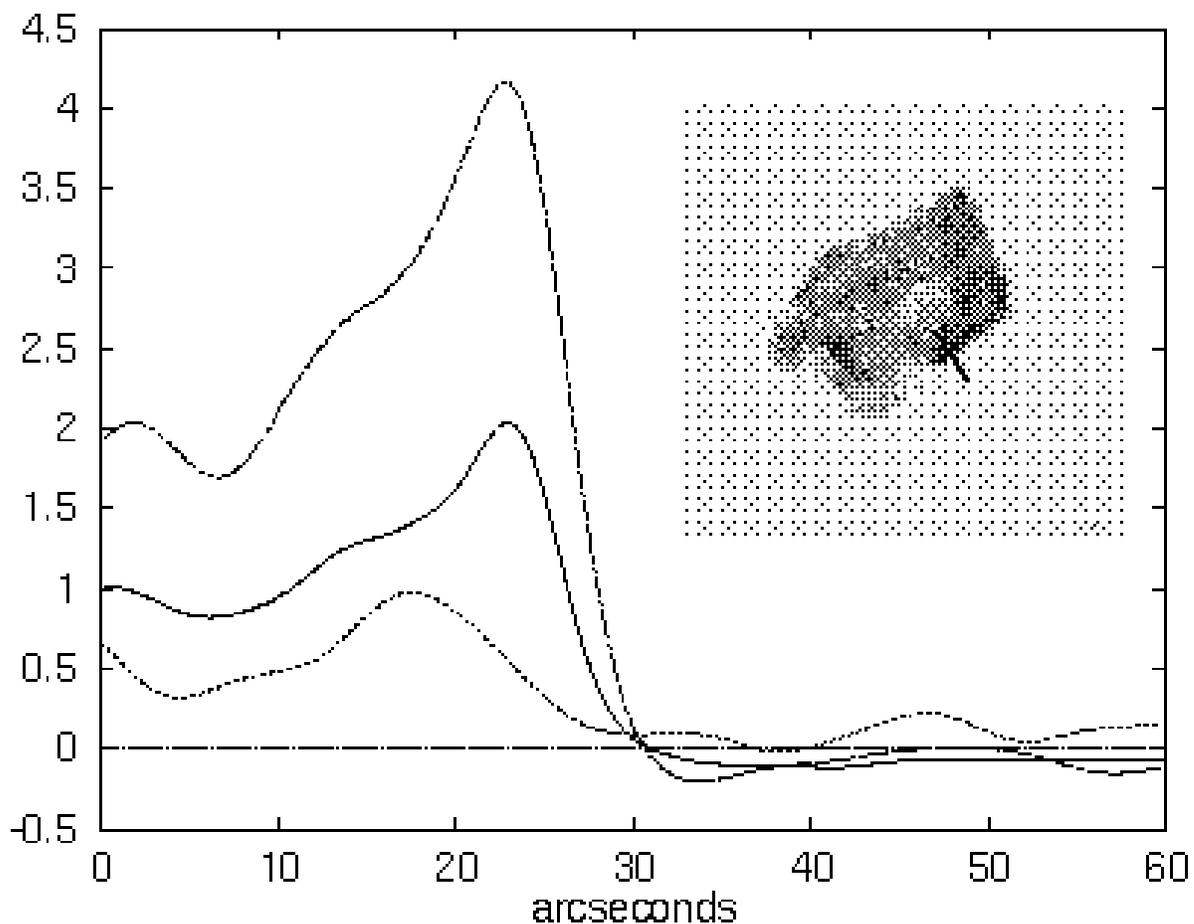}
\caption{Comparison of the steep drop-off at the south-west cliff of 3C 397. 
Dashed line 20 cm, solid line, 6 cm, dotted line X-ray emission. Vertical axis 
is relative scale.\label{cliff}}    
\end{figure}

The close correspondence of X-ray and radio edges can be seen in the
profiles in Figure \ref{cliff}.  Within the limits of our resolution,
it appears that the edges are in the same location, confirming that
electrons are heated and accelerated in the same locations to within
better than 5\% of the remnant radius of 5.8 pc.  The X-ray peak
appears to be shifted inward by about one resolution element (6\arcsec), which 
may reflect slower heating processes, such as
Coulomb equilibration between electrons and ions rather than
instantaneous post-shock equilibration driven by some kind
of plasma instabilities.

\begin{figure}
\plottwo{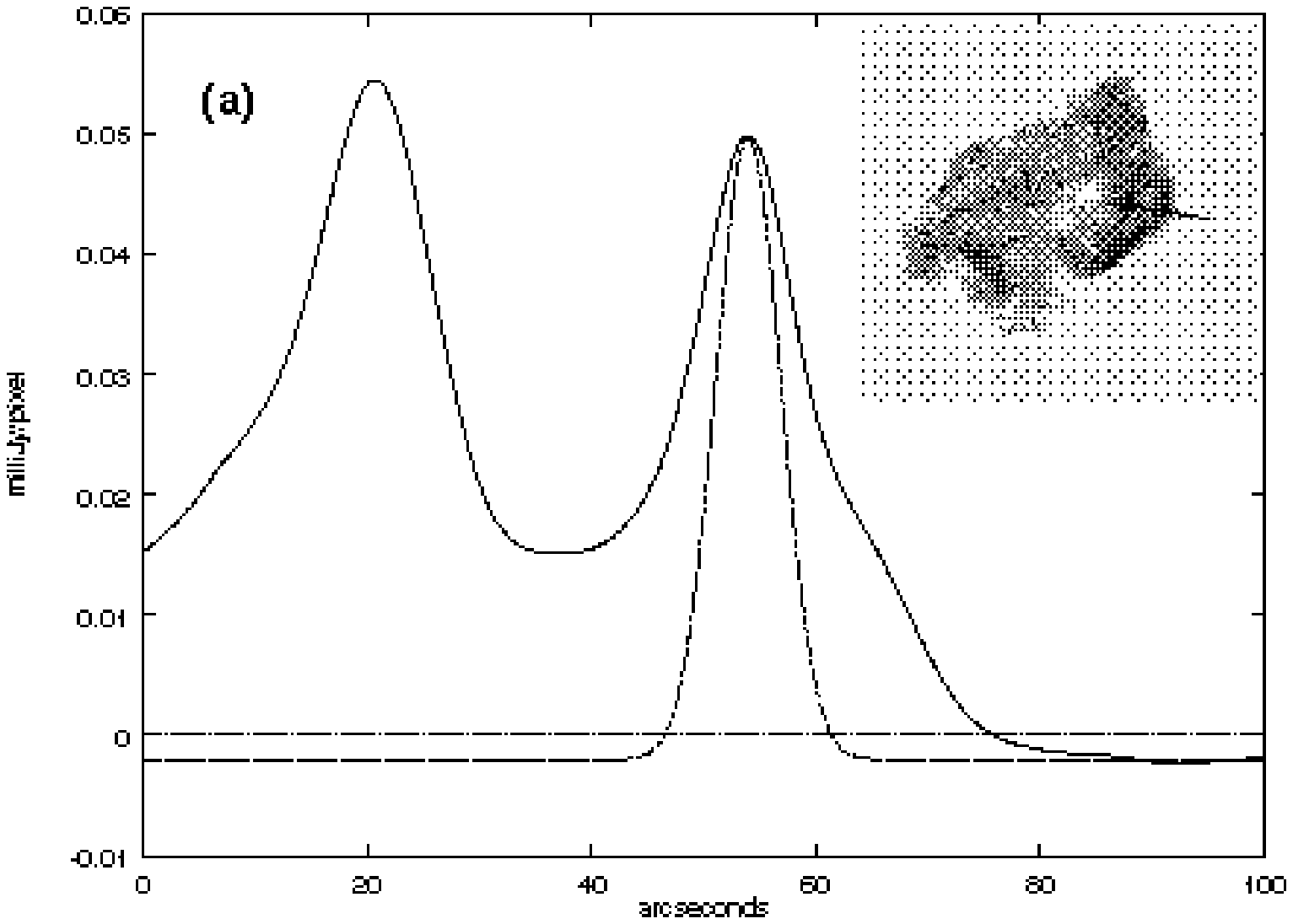}{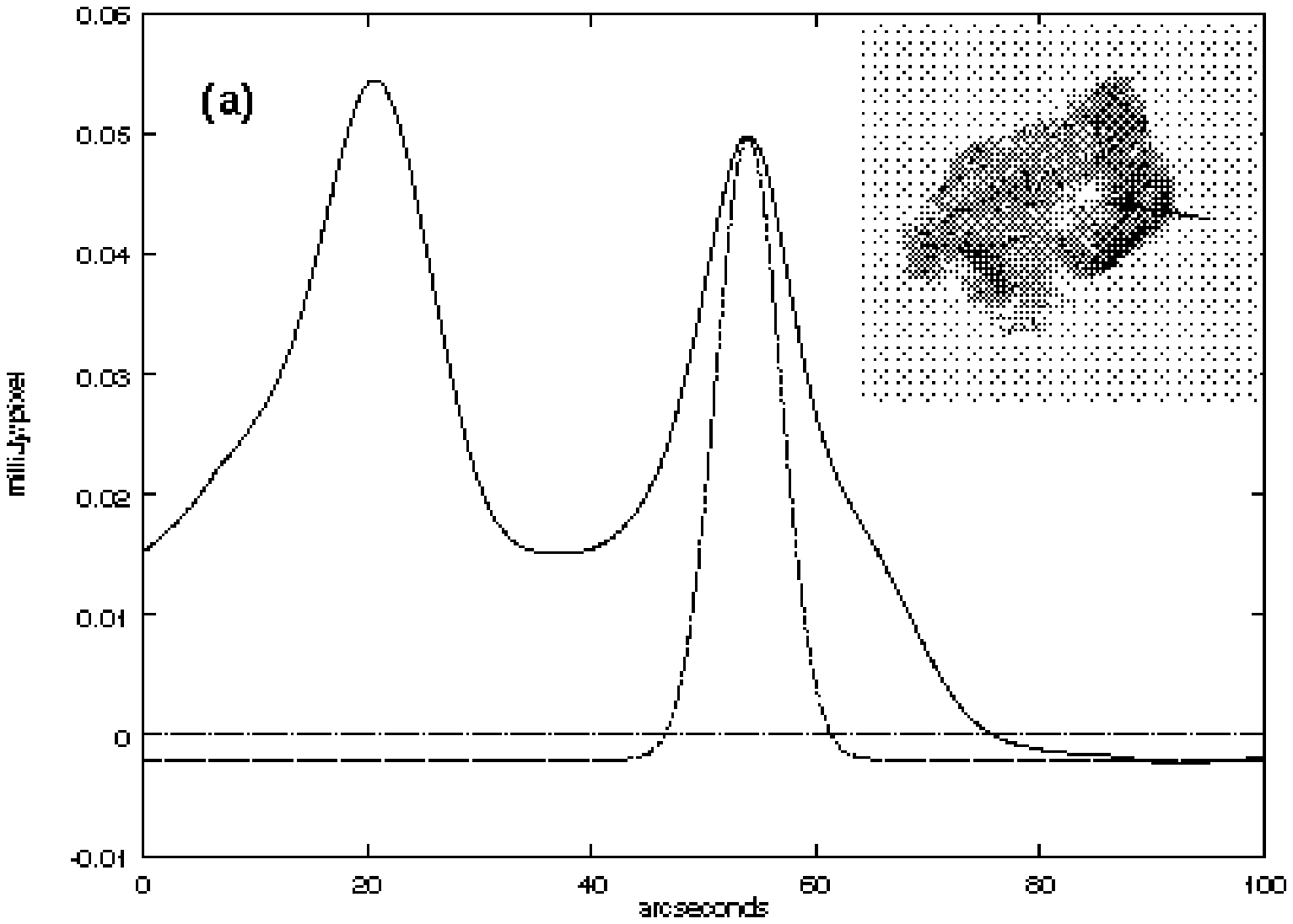}
\caption{Profiles of 3C 397's brightness distribution through the
southwest edge of the remnant showing the possibly resolved drop off the
edge of the remnant. Dashed lines indicates the size of the beam. a) 20 cm image; (b) 6 cm image.\label{shelf}}

\end{figure}

The diffusion of shock-accelerated electrons ahead of the shock should
result in a ``halo'' of faint synchrotron emission beyond the shock
itself (\cite{Achterberg1994} 1994).  Unresolved radio edges can put
limits on the electron diffusion coefficient by requiring any such
halo to be narrower than the diffusion length of radio-emitting
electrons.  The edge of the remnant is most sharply defined across the
south-west drop-off. We examine profiles in two places -- a sharp
cliff at position angle (PA) $\sim $130\arcdeg \ and an equally well defined edge with
a broad shelf at PA $\sim $100\arcdeg \ (see Figure \ref{shelf}).  (The
position angle is measured with respect to the center of the remnant.)
While the beam in its largest dimension is 5.9\arcsec at 20 cm, since the edge
is featureless we assume it to be unresolved, and deduce a nominal
resolution of 7.5\arcsec~at 20 cm and 6.7\arcsec~at 6 cm. The
scattering ahead of the shock is related to the angle between the
shock normal and the external magnetic field $\theta_{\rm Bn}$. \ We
obtain an upper limit on the relativistic electron mean free path
$\lambda_e$ from Eq.~(26) in \cite{Achterberg1994} (1994):
$$\lambda_e~\cos^2(\theta_{\rm Bn})\approx 4\times10^{14}~
\left( \frac{U_s}{2500~\rm{km~s^{-1}}} \right)
\left(\frac{\Delta_{\frac{1}{2}}('')~d(\rm{kpc})}{a} \right)~\rm{cm} $$
where $U_s$ is the shock velocity,~$\Delta_{\frac{1}{2}}$ is the half width at half maximum
measured from the inflection point, $d$ is the distance to the object, and
$a$ is a constant on the order of unity. 

We estimate a shock velocity ${U_s}$ of 1600 km s$^{-1}$ (see below),
and measure the width ~$\Delta_{\frac{1}{2}}$ to be 10\arcsec. If the
distance to the remnant is 10 kpc, we calculate an upper limit to the
mean free path $\lambda_e\cos^2(\theta_{\rm Bn})$ of about
$3\times10^{16}$ cm (0.01 pc).  Unless this part of the shock front
happens to have a special geometry ($\theta_{\rm Bn} \sim 1$, a highly
perpendicular shock), we infer a value for $\lambda_e$ between typical
ISM mean free paths of 0.2 pc and the upper limits found by
\cite{Achterberg1994} for Tycho's and Kepler's remnants of $(1 - 4)
\times 10^{-4}$ pc, indicating an intermediate strength of wave
turbulence ahead of the shock. This limit will not apply if the upstream 
magnetic field is very close to the line of sight, so that synchrotron 
emission from a halo would
be suppressed.  However, our dynamic range of almost 20 (peak edge
brightness to 3$\sigma$) means that the upstream magnetic field would
need to be within a few degrees of the line of sight to make a halo
undetectable by these observations.

\subsection{Polarization and Rotation Measure}

No polarization was detected at 20 cm. From the observed rms of 0.11
mJy beam$^{-1}$ in Stokes $Q$ and $U$ images, we deduce a limit on polarized
intensity $\sqrt{Q^2+U^2}$ of 0.33 mJy beam$^{-1}$, 
corresponding roughly to $3\sigma$.
  Lack of detectable polarization could be due to several
factors including Faraday depolarization in the source, bandwidth
depolarization (large differential Faraday rotation across the
bandwidth), and a highly disordered magnetic field in the source.
Since we did detect some polarization at 6 cm, we suspect a
combination of bandwidth depolarization and internal Faraday
depolarization.

The polarization position angles we do observe (Figure \ref{pola})
appear smoothly varying; some gaps appear between regions of
quite different position angles, suggesting beam depolarization
is causing the gaps.  

Though we could not use 20 cm observations in combination with 6 cm to
remove foreground Faraday rotation, we do have two independent
datasets at 6 cm, at the two IF's separated by 50 MHz, at 4836 and
4886 MHz.  A large Faraday rotation (hundreds of rad m$^{-2}$) could
produce detectable rotation between these two frequencies.  We
attempted to to determine the mean Faraday rotation angle between the
two IFs, since the amount of rotation was too small to yield
significant values in each beam area.  By subtracting the polarization
angle images made from the two IF datasets, a mean rotation of
1.2$^\circ$ was found, implying a rotation measure of around +290 rad
m$^{-2}$. This rotation measure would rotate position angles by about $44^\circ$
between our two 20-cm frequencies of 1446 and 1496 MHz, but over a
6.25 MHz bandwidth by only about 6\arcdeg.  (Our choice of this narrow
bandwidth was made specifically to reduce bandwidth depolarization.)

The remaining possibility to explain the lack of polarization at 20
cm is internal Faraday depolarization.  If the remnant were homogeneous 
with an electron density $n_e$ cm$^{-3}$ and line-of-sight magnetic field 
component $B_\parallel$ gauss over a path length through the source of $L$
pc, the rotation measure would be $8.12 \times 10^5 n_e B_\parallel L$ rad m$^{-2}$
(\cite{Spitzer1978} \ 1978).  For a source path of about 10 pc, and an
assumed $B_\parallel$ of 10 $\mu$gauss, we require an electron density
of 0.7 cm$^{-3}$ to obtain a back-to-front rotation of $\pi$ radians.
If $B_\parallel$ changes sign randomly in cells of length $\ell$, this
estimate rises by $(L/\ell)^{1/2}.$  We conclude that a density
of a few cm$^{-3}$ is adequate to explain the lack of 20-cm polarization.
(The back-to-front rotation will be less at 6 cm by an order of
magnitude.)

\subsection{Spectral Index}


We do not believe physical spectral index variations can be separated
from those induced by the lack of short spacings. Because
interferometers cannot detect flux smoothly distributed on larger
angular scales than those corresponding to their shortest baselines,
they rely on single-dish measurements. However, single-dish
observations of 3C 397 are grossly inconsistent with one another.
Figure \ref{ford} shows the integrated radio spectrum of 3C 397, as
measured by various single dishes. From the scatter, it is clear that
the mean spectral index is not well determined.

We have tried three different methods for studying spectral-index
variations: straightforwardly examining maps of
$\log(S_2/S_1)/\log(\nu_2/\nu_1)$, subtracting the
images with different relative scalings, and using a statistical
method, finding the best-fit line to a plot of many pixels' $S_{6}$
values vs.~their $S_{20}$ values.  All of these procedures were applied
both to CLEAN and to VTESS images.  All of these showed small
variations, but inconsistently with one another, and inconsistent with
the earlier work of \cite{AR1993} (1993), who used the statistical
approach with a somewhat smaller dataset.  The variations we found
implied variations in the 6-cm intensity of amounts comparable to the
uncertainties in missing-flux reconstruction and to the differences
between reconstruction algorithms.  While there may be physical
spectral-index variations in 3C 397 between 20 and 6 cm, our evidence
is also consistent with an artificial dispersion in spectral-index
values of about $\Delta \alpha \sim 0.2$.

In fact, there are no obvious theoretical reasons to expect dramatic
spectral-index variations over the relatively small range in
frequencies of a factor of 3, corresponding to a factor of 1.7 in
electron energies.  Test-particle shock acceleration theory (e.g.,
\cite{Blandford1987} 1987) shows that unless shock Mach numbers are
below 5 or so, the spectral index of synchrotron photons will vary by
less than 0.1 from its asymptotic strong-shock value of --0.5.  The
blast waves in young remnants are expected to have shock waves with
Mach numbers of 50 or higher.  Nonlinear spectral calculations of
electron acceleration in modified shocks (\cite{ER91} 1991) do predict
curved electron spectra flattening to higher energy, by about 0.5 in
electron energy index (0.25 in emitted photon spectral index) over a
range of 100 in electron energies.  Regions of stronger magnetic field
would then be exhibiting lower-energy electrons, with a steeper
spectrum, at a fixed observing frequency ($\nu_{\rm obs} \propto E^2
B$), but to obtain a difference of 0.1 in the photon index would
require magnetic-field variations of factors of order 100.  This seems
very unlikely.  Furthermore, this idea would predict a correlation of
brighter regions with steeper spectra, which is not seen.

Future progress in spectral-index studies of supernova remnants will
require longer frequency baselines and the addition of well-understood
single-dish data to interferometer data.  The uncertainties in typical
single-dish observations mean that such an attempt is likely to be
quite difficult.  We are forced to conclude that the question of
spectral-index variations in 3C 397 in particular is unsettled at this
time.

\subsection{Dynamical Inferences}

  
Even though the {\it PSPC} spectral information is meager, we can make a few
deductions.  The various thermal models described above, though quite
different physically, give roughly the same integrated X-ray flux
between 0.4 and 2 keV, corrected for absorption: $S_x \sim (5 - 10)
\times 10^{-11}$ erg cm$^{-2}$ s$^{-1}$, for an X-ray luminosity in
the ROSAT band of about $10^{36}$ erg s$^{-1}$.  Furthermore, unless
nonequilibrium and abundance effects are very large, the mean
emissivity $\epsilon$ of the X-ray-emitting plasma should not differ
by more than a factor of a few from the mean value for an equilibrium,
cosmic-abundance plasma around 1 keV of about $10^{-23}$ erg cm$^3$
s$^{-1}$.  If we take the mean remnant radius to be $2\arcmin \sim
5.8$ pc, the implied number density in the interior is about
$$n \sim \left( {3 S_x \over {\epsilon f}} \left({d^2 \over R^3}\right) 
            \right)^{1/2} \sim 4 \ {\rm cm}^{-3}$$
where $d$ is the source distance, $R$ the mean radius, and $f$ the
filling factor of the X-ray emitting gas.  An interior density of
about 4 cm$^{-3}$ is consistent with the estimate from radio
depolarization, and implies an upstream density $n_0$ of about $1$
cm$^{-3}$.  Because of the relatively weak dependence on $\epsilon$
and $S_x$, this estimate is unlikely to be very far wrong.  It implies
a swept-up mass of about 29 $M_\odot$, suggesting that unless the
supernova progenitor was very massive, 3C 397 should be well into
transition to Sedov evolution.

If we assume Sedov dynamics and an explosion energy of $10^{51}$ erg,
the observed remnant radius gives an age of 1400 $n_0^{1/2}$ yr.
However, this estimate varies as $d^{9/4}$, so increasing the distance
to 12 kpc would increase the age by 50\%.  The swept-up mass varies as
$d^{5/2}$ as well.  Then the shock velocity now is $1600 \ n_0^{-1/2}
(d/{10 \ {\rm kpc}})^{-5/4}$ km s$^{-1}$.  The implied post-shock
temperature is 3 keV, but if ions and electrons do not equilibrate
immediately, the electron temperature could be lower.  It is
likely that 3C 397 has not fully settled into the Sedov phase, so
that the expansion has been faster and the age lower than this 
estimate.  This also would result in a higher shock speed at present.

The high column density required by the fits, of $1.3 \times 10^{22}$
cm$^{-2}$ or higher, implies a visual extinction of $A_V \gapprox
7^m$, using the empirical optical extinction/column density relation
of \cite{Gorenstein1975} (1975).  This would be more than sufficient
to completely obscure 3C 397 at optical wavelengths, and would remove
any hope of finding a record of 3C 397's supernova among the
historical supernovae.

With an approximate X-ray flux and temperature we can estimate a
distance for 3C 397 by assuming it is in the Sedov phase (Kassim et
al.~1994): 
$$D_s({\rm kpc}) = 8.7 \times 10^6 \epsilon^{0.4}_0 P(\Delta
E, T)^{0.2} \theta^{-0.6} F_{x0}^{-0.2} T^{-0.4}$$
 where $\epsilon_0$
is the initial energy of the SN, $P(\Delta E, T)$ is the thermal X-ray
emissivity in the ROSAT bandpass, $\theta$ is the angular diameter of
the SNR in arcminutes, $F_{x0}$ is the observed X-ray flux in the
ROSAT bandpass, corrected for absorption, and $T$ is the post-shock
temperature in K.  This formulation assumes that the remnant is fully
in the Sedov phase and that electron and ion temperatures are equal.
The weak dependence on the emissivity means that nonequilibrium
effects do not make much difference.  For our values of $P(\Delta E,
T) \sim 10^{-23}$ erg cm$^3$ s$^{-1}$, $\theta = 4^\prime$, $F_{x0} =
(5 - 10) \times 10^{-11}$ erg cm$^{-2}$ s$^{-1}$, $T = 2 \times 10^7$
K ($kT$ = 1.8 keV), and assuming a canonical explosion energy of
$10^{51}$ erg, we get a distance estimate between 11 and 13 kpc --
quite consistent, given the idealized assumptions, with that obtained
from H I observations. Certainly ASCA data will allow a more detailed 
and accurate fit (\cite{Samar} 1999).


\subsection{The Nature of the Unresolved X-ray Source}

The X-ray hot spot has no counterpart at radio wavelengths, in total
intensity or polarization. It is also not apparent in the
hardness-ratio map.  However, since the feature represents less than
7\% of the total X-ray emission, a spectral anomaly could easily be
lost in the narrow ROSAT band, especially since absorption effectively
limits the band to only a factor of 4 in photon energy.  We consider
various explanations for this feature.

We can roughly estimate the luminosity of the feature by assuming a
spectral shape of the best-fitting Raymond-Smith spectrum described
above ($kT = 1.8$ keV): we obtain about $4 \times 10^{34}$ erg
s$^{-1}$ at a distance of 10 kpc, after correcting for absorption. 
The least interesting explanation would involve a density
inhomogeneity, perhaps in relatively unmixed ejecta where the
synchrotron emissivity might be expected to be small (since the ejecta carry only the highly diluted stellar magnetic field).

However, the unresolved core could certainly be a remnant neutron
star, seen either in thermal emission or in pulses.  The thermal
luminosity of a neutron star with $T = 2 \times 10^6$ K ($kT = 0.17$
keV) and a radius of 15 km would be several times 10$^{34}$ erg
s$^{-1}$.  However, this very soft thermal emission would require a
much greater luminosity to provide the observed flux of the hot spot
behind $N_H \sim 10^{22}$ cm$^{-2}$.  While quantitative statements
are difficult in view of all the uncertainties, it is likely that
thermal emission from a neutron star would be too soft to produce the
observed unresolved spot.  In any case, we would still need to
attribute the extended portion of the feature to some density enhancement in the thermal gas.

A mini-plerion is unlikely due to the various arguments made above, 
but might still be possible if the population of low-energy
radio-emitting electrons is somehow depleted.  Since it is not clear
where the Crab Nebula's radio-emitting electron population comes from
(Kennel and Coroniti 1984), this might not be too far-fetched.  This
explanation could take care of the resolved part of the feature as
well.  The limits on direct pulsations are weak, but we have observed
3C 397 with XTE and the examination of this dataset could either
produce, or put much stronger limits on, X-ray pulses.

The low luminosity rules out a standard X-ray binary with
characteristic temperature of order a keV or higher.  To reach the
luminosity of even relatively weak low-mass X-ray binaries of around
$10^{36}$ erg s$^{-1}$ would require that the unresolved core be at a
distance of only 2 kpc, superposed by chance against 3C 397 (and on
top of the resolved feature).  No star or other optical feature can be
seen at the hot spot position on Palomar Sky Survey images.  However,
an anomalously soft-spectrum X-ray binary would need to be more
luminous to produce the observed flux.  Finally, the feature could be
a chance background quasar at 1000 Mpc, with a typical active galactic nucleus X-ray
luminosity of order $10^{44}$ erg s$^{-1}$, as long as any associated
radio emission was faint enough to be hidden in the remnant's interior  
radio emission.

The source is made to order for observations with the Chandra
Observatory (AXAF).  Its small angular size will allow optimal use of
Chandra's high angular resolution, and there should be enough spectral
information in Chandra's much larger X-ray bandpass to separate the
very distinct spectral predictions of the different explanations: very
soft blackbody spectrum for a cooling neutron star; optically thin
plasma for a density bump; nonthermal, featureless power-law for a
plerion.  Of course Chandra's CCDs will also do a good job obtaining
spatially resolved spectra for a better understanding of the entire
remnant.

\section{Conclusion}
\label{sec:Conclusions}

Our radio and X-ray study of 3C 397 has at least settled its nature as
a thermal shell remnant without an obvious plerionic component, but it 
has produced several more interesting questions.  The relatively good
correspondence between radio and X-ray morphologies puts 3C 397 in a
sequence from the historical shells such as Tycho and Cas A, with
excellent correspondences, to the still radio-bright but clearly more
evolved objects like W49 B in which the X-ray morphology bears little
relation to the radio.  This correspondence provides more evidence for
3C 397's youth.  The {\it PSPC} spectrum is poorly fit by any reasonable
model, and may require a composite model including ejecta.  Such a
study demands spectral data of at least the quality of ASCA.

We find weak mean radio polarization of $1.5 \pm 0.1$\% at 6 cm,
with polarized flux not well correlated with total radio intensity
or with X-rays (and no polarization feature coincident with the X-ray
hot spot).  That low fraction, and our failure to detect any polarization
at 20 cm, are consistent with internal Faraday depolarization by
material with a mean density of about 4 cm$^{-3}$ as derived from the
X-ray fitting.

In agreement with past studies, we find relatively small radio
spectral-index variations over the narrow range in wavelengths from 20
to 6 cm.  It should be remembered that this range covers only a    
factor of 1.7 in electron energies, and large spectral-index
variations would cause the integrated spectrum to show significant
concave curvature.  The poor state of total-flux observations of 3C
397 cannot, however, rule out such a situation.

The X-ray data in the 0.5 -- 2 keV {\it PSPC} band do not allow detailed
spectral fits, but can be roughly described by an equilibrium plasma
with a temperature of about 1.8 keV.  The observed X-ray flux implies
the mean density of 4 cm$^{-3}$ just mentioned.  Simple dynamical
estimates then give a remnant age of order $10^3$ yr, and a current
shock velocity of about 1600 km s$^{-1}$.

The mysterious central X-ray source must be studied with an instrument
with sensitivity at higher energies, as well as higher spatial
resolution.  The Chandra X-ray Observatory will be an ideal choice.
X-ray observations to higher energies, along with more sophisticated
spectral modeling, will be the best way to improve our understanding
of 3C 397 and its evolutionary stage. 

\acknowledgments

We acknowledge with appreciation extensive discussions with
K.~Borkowski and L.~Rudnick. We are also grateful for useful comments by Namir Kassim. This research was supported by NASA
grants NAG5-2212, NAG5-2844, and NGT5-65.

Our research made use of the following online services: NASA's
Astrophysics Data System Abstract Service, NASA's SkyView facility
(http://skyview.gsfc.nasa.gov) located at NASA Goddard Space Flight
Center and SIMBAD at Centre de Donn\'ees astronomiques de Strasbourg (US
mirror http://simbad.harvard.edu/Simbad).

\newpage

\clearpage

\newpage 


\begin{deluxetable}{lcc}
\footnotesize
\tablecaption{Gross Properties of 3C 397 (G41.1--0.3) \label{facts}}
\tablewidth{0pt}
\tablehead{}
\startdata
$\alpha,~\delta~(J2000)$&19$^h$ 7$^m$ 33$^s$
&7\arcdeg~8\arcmin~16\arcsec\nl
(VLA pointing position)\nl
Galactic $l,b$ &41\arcdeg~07\arcmin~0\arcsec 
&--00\arcdeg~18\arcmin~27\arcsec~~\nl
Spectral Index	&0.48$^a$\nl
Integrated Flux at 1 GHz & 22 Jy$^a$\nl
Distance &$\sim$10 kpc\nl
Mean radius & $\sim$2\arcmin~at 10 kpc = 6 pc\nl
\tablerefs{(a) Green 1998}

\enddata
\end{deluxetable}                        
\clearpage

\begin{deluxetable}{lcccc}
\footnotesize
\tablecaption{VLA Observations of 3C 397 \label{observations}}
\tablewidth{0pt}
\tablehead{
\colhead{Date}&\colhead{Configuration}&\colhead{Frequency 
(MHz)}&\colhead{Bandwidth (MHz)}&\colhead{Integration Time (m)}}
\startdata
1991 Apr 12   &D& 4847& 25.00 & 52\nl
		&& 1468& 6.25 & 65 \nl
1990 Oct 21   &C& 4847& 25.00 & 59\nl
		&& 1468 & 6.25& 60\nl
1990 Aug 21,22&B& 1468& 6.25  & 60\nl
\enddata
\end{deluxetable}

\begin{deluxetable}{lcccc}
\footnotesize
\tablecaption{Summary of VLA Total Intensity Images of 3C 397 \label{summary}}
\tablewidth{0pt}
\tablehead{\colhead{Wavelength}&\colhead{Resolution}&\colhead{Detected 
SNR Emission}&\colhead{rms}}
\startdata
\cutinhead{CLEAN}
6 cm  & 6.4 x 5.6\arcsec~PA -59\arcdeg& 5.6 Jy & 0.41 mJy beam$^{-1}$  \nl

20 cm & 6.9 x 6.6\arcsec~PA 84\arcdeg& 13.8 Jy & 0.51 mJy beam$^{-1}$ \nl
\cutinhead{VTESS} 
6 cm & 6.9 x 5.5\arcsec~PA -74\arcdeg& 6.0 Jy & 0.19 mJy beam$^{-1}$\nl
20 cm& 6.1 x 5.7\arcsec~PA 85\arcdeg& 14.1 Jy & 0.29 mJy beam$^{-1}$\nl
\enddata
\end{deluxetable}    
\clearpage

\begin{deluxetable}{lcc}
\footnotesize
\tablecaption{Noise (rms fluctuations)\label{noise}}
\tablewidth{0pt}
\tablehead{\colhead{Method}&\colhead{6 cm (mJy beam$^{-1}$)}&\colhead{20 cm (mJy 
beam$^{-1}$)}}
\startdata
Theoretical 	& 0.05	& 0.10\nl
Stokes $V$	& 0.05	& 0.11\nl
CLEAN		& 0.41	& 0.51\nl
VTESS		& 0.19	& 0.29\nl
\enddata
\end{deluxetable}    

\begin{deluxetable}{lcc}
\footnotesize
\tablecaption{ROSAT {\it HRI} and {\it PSPC} Observations of 3C 397 \label{rosat}}
\tablewidth{0pt}
\tablehead{
\colhead{Parameter}&\colhead{HRI}&\colhead{PSPC}}
\startdata
Cycle & A05& A03\nl
Date & 1994 Oct 14-21& 1992 Sept 28\nl
Pointing center $\alpha$ & 19$^h$ 07$^m$ 33\fs6  \nl
$\beta$ (J2000) &  7\arcdeg~08\arcmin~24\farcs0 \nl
Exposure (ks)& 6.2 & 4.2 \nl
Counts & 1200 & 2700 \nl
Effective Resolution & $\sim$6\arcsec FWHM & $\sim$30\arcsec\nl
Field of view & 38\arcmin~x 38\arcmin & 2\arcdeg (diameter) \nl 
Energy sensitivity &0.1-2.0 keV &0.1-2.4 keV\nl
\enddata
\end{deluxetable}    
\clearpage





\begin{thebibliography}{DUM}  


\bibitem[Achterberg et al.]{Achterberg1994} Achterberg, R. D., Blandford, R. D., \& Reynolds, S. P. 1994, \aj, 281,220

 

\bibitem[Anderson \& Rudnick]{AR1993}Anderson. M. C., \& Rudnick, L. 1993, \apj, 408, 514

\bibitem[Becker et al.]{Becker1985} Becker, R. H., Markert, T., \& Donahue, M. 1985, \apj, 296, 461


\bibitem[Blandford \& Eichler]{Blandford1987}Blandford, R. D., \& Eichler,
D. 1987, Phys.Rep., 154, 1

\bibitem[Caswell et al.]{Caswell1975a}Caswell, J. L., Murray, J. D., Roger, R. S., Cole, D. J. \& Cooke, D. J. 1975, A\&A, 45, 239

\bibitem[Caswell et al.]{Caswell1982} 
Caswell, J. L., Haynes, R. F., Milne, D. K. \& Wellington, K. J. 1982, 
\mnras, 200, 1143

\bibitem[Cersosimo \& Magnani]{Cersosimo1990} Cersosimo, J. C. \& 
Magnani, L. 1990, \aap, 239, 287 

    
   


\bibitem[Downes et al.]{Downes1980}Downes, D., Wilson, T. L., Bieging, J. \& 
Wink, J. 1980, \aaps, 40, 370

\bibitem[Ellison \& Reynolds]{ER91}Ellison, D. C., \& Reynolds, S. P.
1991, \apj, 382, 242

\bibitem[Elvis, et al.]{Elvis1992} 
Elvis, M., Plummer, D., Schachter, J., \& Fabbiano, G. 1992, \apjs, 80, 
257 



\bibitem[Gorenstein]{Gorenstein1975}Gorenstein, P. 1975, \apj~198, 95

\bibitem[Green]{Green1998} Green D.A. 1998, A Catalogue of Galactic
Supernova Remnants (1998 September version), Mullard Radio Astronomy 
Observatory, Cambridge, United Kingdom (available on the World-Wide-Web at 
``http://www.mrao.cam.ac.uk/surveys/snrs/'').



\bibitem[Kassim]{Kassim1992} Kassim, N. E. 1992, \aj, 103, 
943 


\bibitem[Kassim]{Kassim1989A} Kassim, N. E. 1989a, \apj, 347, 
915 

\bibitem[Kassim]{Kassim1989B} Kassim, N. E. 1989b, \apjs, 71, 
799 


\bibitem[Kassim et al.]{Kassim1994} Kassim, 
N. E., Hertz, P. , Van Dyk, S. D. \& Weiler, K. W. 1994, \apjl, 427, L95 


\bibitem[Kennel \& Coroniti]{Kennel1984} Kennel, C. F. \& 
Coroniti, F. V. 1984, \apj, 283, 710 


\bibitem[Koyama et al.]{Koyama1995} Koyama, K., Petre, R., Gotthelf, E. V.,
Hwang, U., Matsura, M., Ozaki, M., Holt, \& S. S. 1995, Nature 378, 255



\bibitem[Paper I]{I} Moffett, D. A. \& 
Reynolds, S. P. 1994, \apj, 425, 668 

\bibitem[Paper II]{II} Moffett, D. A. \& 
Reynolds, S. P. 1994, \apj, 437, 705 


\bibitem[Pohl \& Esposito]{Pohl1998} Pohl, M.  \& Esposito, 
J. A. 1998, \apj, 507, 327

\bibitem[Pye et al.]{Pye1984} Pye, J.P., Becker, R.H., Seward, F.D., 
\& Thomas, N. 1984, MNRAS, 207, 649

\bibitem[Reich et al.]{Reich1990} Reich, W., \& Reich, P., \& F\"{u}rst, E. 1990, \aap~Suppl. Ser. 83, 539 

\bibitem[Reynolds et al.]{Reynolds1994} 
Reynolds, S. P., Lyutikov, M., Blandford, R. D. \& Seward, F. D. 1994, 
\mnras, 271, L1 


\bibitem[Rho \& Petre]{Rho1997}Rho, J., \& Petre, R. 1997, \apj, 484, 828 


\bibitem[Rho \& Petre]{Rho1998} Rho, J. \& Petre, R. 
1998, \apjl, 503, L167 

\bibitem[Safi-Harb et al.]{Samar} Safi-Harb, S., Petre, R., Arnaud, K.A., Borkowski, K., Reynolds, S.P., Dyer, K.K., Keohane, J.W., 1999 in preparation

\bibitem[Saken, Fesen, \& Shull]{Saken1992} Saken, J. M., 
Fesen, R. A. \& Shull, J. M. 1992, \apjs, 81, 715 

\bibitem[Spitzer]{Spitzer1978}Spitzer, L. 1978, Physical Processes
in the Interstellar Medium, p.~66

\bibitem[Tanimori et al.]{Tanimori1998} Tanimori, T., et al. 
1998, \apjl, 497, L25



\bibitem[Vasisht et al.]{Vasisht1996} Vasisht, G., Aoki, T., 
Dotani, T., Kulkarni, S. R. \& Nagase, F. 1996, \apjl, 456, L59 
 


\end{thebibliography}
\end{document}